# Adsorption and binding dynamics of graphene-supported phospholipid membranes using the QCM-D technique

D. A. Melendrez[a], T. Jowitt[b], M. Iliut[a], A.F. Verre[a], S. Goodwin[c] and A. Vijayaraghavan[†a,c]

We report on the adsorption dynamics of phospholipid membranes on graphene-coated substrates using the quartz crystal microbalance with dissipation monitoring (QCM-D) technique. We compare the lipid vescle interaction and membranne formation on gold and silicon dioxide QCM crystal surfaces with their graphene oxide (GO) and reduced (r)GO coated counterparts, and report on the different lipid structures obtained. We establish graphene derivative coatings as support surfaces with tuneable hydrophobicity for the formation of controllable lipid structures. One structure of interest formed are lipid monolayer membrannes which were formed on rGO, which are otherwise challenging to produce. We also demonstrate and monitor biotin-avidin binding on such a membranne, which will then serve as a platform for a wide range of biosensing applications. The QCM-D technique could be extended to both fundamental studies and applications of other covalent and non-covalent interactions in 2-dimensional materials.

[a.] School of Materials and National Graphene Institute, University of Manchester, Manchester M13 9PL UK.
[b.] Biomolecular Analysis Core Facility, Faculty of Life Sciences, University of Manchester, M13 9PL UK.
[c.] School of Computer Science, University of Manchester, Manchester M13 9PL UK.
† Corresponding author email: aravind@manchester.ac.uk

## Introduction

Graphene is a versatile 2-dimensional (2-d) nanomaterial which has attracted particular attention of scientists in the field of biological sensors [1] and biotechnology [2]. Graphene's large surface area [3], biocompatibility [4] and ease of functionalization [5], [6] provide opportunities for biomedical applications [7], [8]. Graphene derivatives can solubilize and bind drug molecules and thus have the potential to be drug delivery vehicles [9]. Such properties can be exploited to increase the sensitivity of biosensors [10] and may act as a supporting platform for the construction of biological detection arrays [11], [12]. Graphene oxide (GO) has played an important role in the development of electrochemical [10], [13] and mass-sensitive sensors [14], [15] and inclusively it has shown strong antibacterial activity [16] by affecting the integrity of the cell membrane [17], which is formed by a continuous lipid bilayer. In GO, the main surface functional groups are hydroxyls and epoxies [18], with carboxylic acids and other keto groups on the edges [19]. GO retains $sp^2$-carbon domains as well as $sp^3$-carbon groups, endowing it with both hydrophobic and hydrophilic domains, respectively [20]. A majority of such functional groups can be removed upon treatment with a reducing agent, such as L-ascorbic acid [21] or hydrazine [22], or thermally [23] to form reduced (r)GO, which is overall significantly more hydrophobic, comparable to pristine graphene.

Exploring different routes for the construction of supported lipid membranes (SLMs) (Scheme 1a), such as supported lipid bilayers (SLBs) or monolayers is pertinent since they are useful platforms for the study of fundamental cell functions and signaling, cell-cell interactions, drug delivery and biosensing [24]. SLMs help to establish a model for the cell membrane and can serve as the key component of biosensor devices. The most common membrane lipids are phospholipids; amphiphilic molecules composed of hydrocarbon chain tails attached to a phosphate head group.

Here we present the study of the adsorption and rupture of small unilamellar vesicles (SUVs) (typically <100 nm) on GO- and rGO-substrates, monitored using the quartz crystal microbalance with dissipation monitoring (QCM-D) technique [25]. Substrates used here were produced by spin coating GO on QCM-D crystals. This coating technique is a fast and reliable way that renders a uniform surface coverage with small quantities of GO dispersion (<100 μL/crystal) and serves as the first step to obtain an rGO film upon thermal reduction of selected substrates. Moreover, both the automation of sample injection and the real-time monitoring capabilities of the QCM-D (Q-Sense Pro, Biolin Scientific, Gothenburg, Sweden) system improve the experimental

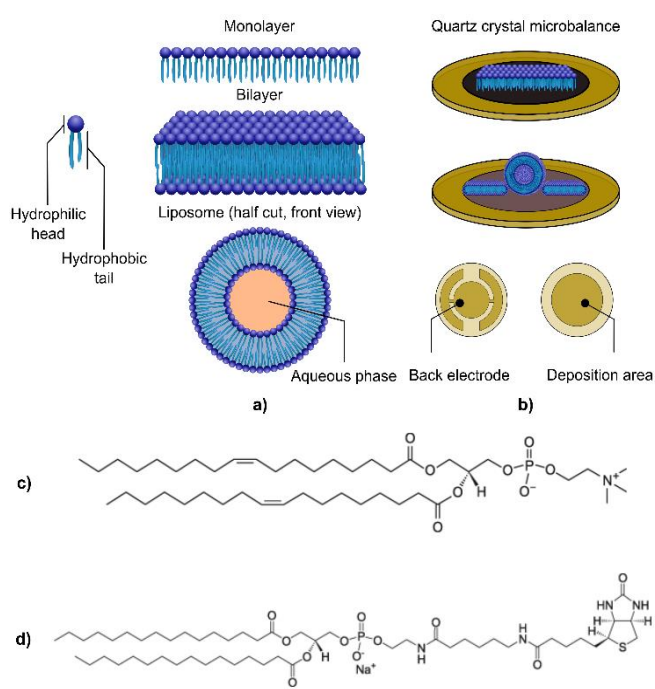

**Scheme 1.** Schematic representation of lipid membranes from the basic lipid unit. a) The three possible lipid structures studied b) Depiction (not drawn to scale) of a lipid monolayer on reduced graphene oxide and intact vesicles supported on graphene oxide on top of a Quartz Crystal Microbalance chip c) DOPC molecular structure d) Biotinyl cap PE molecular structure.

performance and simplifies the interpretation of the dynamics of membrane formation.

The interaction of graphene derivatives and phospholipid membranes using different techniques, such as AFM, ellipsometry and electrochemical cells has been previously reported [26]–[34] and the results are promising in the development of highly versatile biosensors. However, the formation of different lipid structures on commonly used substrates as well as on graphene is still not completely understood. Changing the physiochemical properties and the topology of the substrate that acts as a mediator of the interaction between the phospholipids and the surface is key in establishing possible mechanistic scenarios for the adsorption and spreading into specific structures (Scheme 1a-b). On this regard, the different chemistries present in GO and rGO provide certain hydrophilic and hydrophobic degrees and surface net charge which favor the formation of specific lipid structures. The hydrophilicity of SiO$_2$ makes it the ideal surface for SLB formation via vesicle fusion which on plain Au has been difficult to achieve using the same technique. Therefore, finding adequate modifiers to obtain different surface chemistries is crucial in the development of biocompatible platforms. Accordingly, our investigation is primarily focused on the study of biomimetic lipid membranes, namely, those that fully or partially mimic the structure of the naturally occurring containment unit of the cell as well as equivalent lipidic structures with potential biomedical applications.

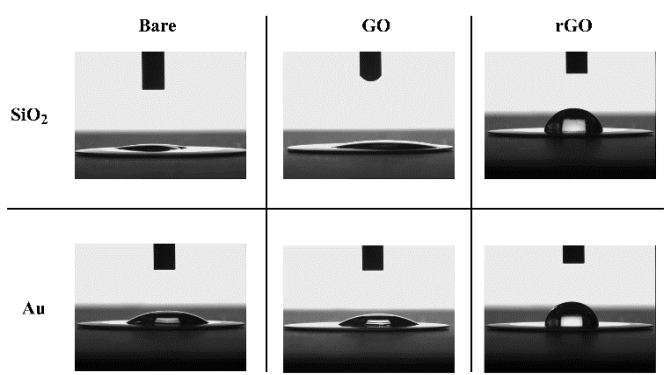

**Fig. 1** Optical images of the drop shape obtained during contact angle measurements on the bare and graphene coated SiO$_2$ and Au chip surfaces.

**Table 1** Mean values of wetting contact angles measured on the bare and graphene coated SiO$_2$ and Au chip surfaces.

|      | SiO$_2$ | Au |
| --- | --- | --- |
| **Bare** | 17.34º ± 2.79º | 41.32º ± 0.83º |
| **GO** | 32.12º ± 3.18º | 38.12º ± 1.43º |
| **rGO** | 88.44º ± 1.40º | 94.74º ± 1.39º |

The formation of lipid membranes on bare and modified SiO$_2$, Au, mica and TiO$_2$ substrates have been previously reported and discussed in depth [35]–[38]. Promoting adsorption of SUVs onto a substrate followed by spontaneous rupture and the formation of a uniform membrane involves control over various parameters such as: vesicle size [37], electrostatic force [36], ionic properties of the buffer solution [39], [40], as well as vesicle-vesicle and vesicle-substrate interactions [41]. The latter will strongly depend on the properties of the substrate therefore varying the surface chemistry will lead to different lipidic conformations which can be monitored in real time with the help of acoustic wave sensors. Specifically, the QCM system has emerged as the primary instrument for ultra-sensitive mass detection ( <1 ng/cm$^2$) [25]. The QCM-D technique has drawn great interest as reported in studies on lipid adsorption kinetics [36], [42], measuring vitamin-protein binding events [43], [44], studying interactions in layer-by-layer graphene-lipid structures [45] and formaldehyde detection using GO as a sensing layer [46]. Modifying the supporting substrate to obtain different lipidic structures is one common approach, e.g. creating a hydrophobic layer through the deposition of -thiol self-assembled monolayers (SAMs) [41], using polymers as cushions [47] or using SAMs of organosulfates and organophosphates [48] have been reported. These modifiers are mostly used to provide specific functional groups that confer hydrophilic or hydrophobic domains, to act as spacers for further protein insertion into the membrane or to change the charge of the substrate.

Graphene's high surface area, biocompatibility, ease of functionalization and electrical properties makes it a great candidate as a surface modifier with similar simplicity of SAMs [47]. On this regard, GO is particularly interesting from its multiple functionalities. A large amount of research is devoted to understand how lipids interact with graphene to form well-defined lipidic structures and is an emerging area of investigation. Recently, Tabaei et al., [49] reported the formation of a lipid monolayer on pristine graphene grown via the chemical vapor deposition (CVD) technique and then transferred to a SiO$_2$ coated QCM-D chip. Then, through the vesicle fusion technique and via a solvent-assisted lipid bilayer method the authors reported the formation of a supported layer of unruptured vesicles and a lipid bilayer on oxidized CVD graphene, respectively. Their findings are in line with previous reports which have highlighted the pivotal role that both the hydrophilic and hydrophobic interactions play in the support of lipid membranes [26], [50]–[52]. One disadvantage of the growth of graphene through the CVD technique and transfer onto a supporting substrate is that it involves a harsh atmosphere. In this regard, the annealing temperature to obtain an rGO film reported in this work is relatively low (180 °C), in contrast to the high temperatures (> 900 °C) reached for thermal reduction [20], [53]. In contrast, using graphene dispersions, like GO, for the formation of thin film coatings offers a fast route to achieve a surface with tunable hydrophobicity through thermal reduction, which in consequence helps to recover most of the properties of pristine graphene. Therefore, we propose a friendly method to obtain graphene-based acoustic wave biosensors.

In this report, we present the adsorption dynamics of zwitterionic lipid vesicles on two traditional substrates, SiO$_2$ and Au, whose hydrophobicity is modified by two graphene derivatives: GO and rGO. We additionally investigated the utility of the formed lipid membranes through a biomolecular binding event, specifically between the biotin-avidin complex which is the strongest known non-covalent biological interaction [54], [55] and is used for the development of robust and highly sensitive assays useful in protein detection [56], [57]. Non-covalent intermolecular interactions involving π-systems are pivotal to the stabilization of proteins, enzyme-drug complexes and functional nanomaterials [5], [58]. One possible route to experimentally accomplish this binding event is by presenting the avidin protein dispersed in buffer to lipid membranes formed from biotinylated lipid vesicles that have been adsorbed, and in some cases ruptured and reorganized on a supporting substrate.

## 2. Experimental methods
### 2.1 Solutions and reagents preparation

GO dispersions were prepared by oxidizing graphite flakes according to a modified Hummers method [59] followed by exfoliation and purification, as described in the electronic supporting information (ESI).

Detailed protocol for the preparation and assembly of phospholipid membranes is presented in the ESI. Briefly, the buffer solution was prepared with 10 mM HEPES (4-(2-hydroxyethyl)-1-piperazineethanesulfonic acid) buffer from powder (Sigma Aldrich), 100 mM NaCl, and 5 mM MgCl$_2$, all then diluted in ultrapure water (18.5 MΩ, MilliQ). The pH was corrected to 7.4, dropwise with a solution of NaOH. This buffer solution was filtered with 0.22 $\mu$m pore-size nylon membranes before each experiment and stored in the fridge for up to two weeks.

To prepare 1 mL of DOPC lipid vesicles 1 mL of 2.5 mg/mL DOPC (1,2-dioleoyl-sn-glycero-3-phosphocholine) lipid (Avanti Polar Lipids) dispersed in chloroform, was dried under a stream of nitrogen before resuspending in 10 Mm HEPES. For the binding studies, 10 % biotinyl cap PE lipid were mixed to the final concentration of 10% total lipid in chloroform before drying. The functionalized lipid was hydrated with the HEPES buffer solution and subsequently extruded more than 23 times, as recommended by Cho's protocol [36], using 50 nm pore-size polycarbonate membranes to obtain vesicles with a diameter size distribution of ∼80-110 nm, analyzed via dynamic light scattering (DLS, Malvern Zetasizer Nano-S) (data not shown).

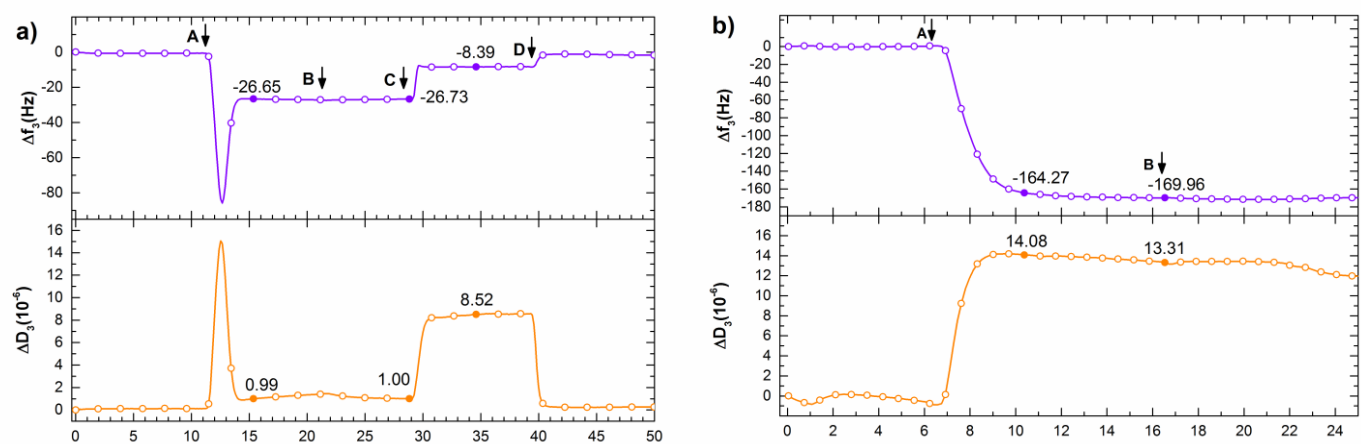

**Fig. 2** Frequency (purple, top) and dissipation (orange, bottom) response from the adsorption of DOPC on bare substrates: **a)** on bare $SiO_2$ and **b)** bare Au. Steps: (**A**) DOPC injection, (**B**) buffer rinse, (**C**) SDS rinse, (**D**) final buffer rinse.

Avidin from egg white (Thermo Fischer Scientific) from a stock concentration of 1mg/mL was diluted to 50 μg/mL. Finally, the extruded lipid was aliquoted with a ratio of 1:10 (lipid:buffer) and stored in the freezer at -20 °C for up to two weeks to ensure vesicle stability.

### 2.2 QCM chips preparation and equipment setup

AT-cut piezoelectric quartz crystals from Q-Sense (Biolin Scientific, Gothenburg, Sweden) with a fundamental frequency of 5 MHz were used in all experiments. $SiO_2$ and Au coated crystals were first immersed in a 2% solution of sodium dodecyl sulfate (SDS) and sonicated in an ultrasonic bath during 25 minutes, rinsed with copious amounts of ultrapure water, then soaked in 99% ethanol for 10 min, then dried under a soft beam of nitrogen gas and immediately placed inside a UV/Ozone system to be treated for 30 min. Crystals were spin coated at 3500 rpm during 2 min (Laurell Tech., WS-650MZ) using the final dilution of GO (0.5 mg/mL).

For the thermal reduction of GO coated crystals to obtain an rGO film, GO coated sensors were placed in the oven at 180 °C under vacuum (Towson + Mercer, EV018) during 20 hours.

The Q-Sense Omega Auto (Q-Sense, Gothenburg, Sweden) system was used to monitor the adsorption kinetics of lipid vesicles on bare and coated sensors and record the frequencies (3rd to the 7th harmonics) and dissipation values. We report the value for the 3rd overtone for both frequency ($\Delta f_{n=3}/3$) and energy dissipation ($\Delta D_{n=3}/3$) components. Values at other overtones can be found in the ESI.

For the wetting properties analysis, a Kruss DSA100 (Hamburg) system was used measure the wetting contact angles (WCAs) for the bare, GO- and rGO-coated QCM chips. A manually controlled syringe was used to cast sessile drops (∼5 μL) of DI water on top of the substrates under study. An ImageJ plugin for contact angle measurement developed by Marco Brugnara [60] was used to compute the WCAs (Table 1) through a manual ellipse/circle fitting method.

## 3. Results and discussion
### 3.1 Membrane formation overview

Figure 1 shows the DI water droplets formed on the surface of the QCM crystals (see ESI for extended wetting analysis). The corresponding contact angles are summarized in Table 1. Preference for intact vesicle adsorption and/or spreading into a mono- or bilayer is influenced by the hydrophobicity or hydrophilicity of the deposition substrate. Removing organic contaminants by means of chemical agents and UV/Ozone treatment is crucial to render a substrate hydrophilic [61]. On this regard, the formation of a SLB requires a hydrophilic substrate, where a critical coverage of adsorbed vesicles must be reached first, since vesicle-substrate interactions usually do not commence individual vesicle rupture [36]. In particular, $SiO_2$ has been the most widespread substrate for the formation of SLBs and the adsorption dynamics is well understood. On the other hand, the hydrophobic vesicle-substrate interaction has been considered as the primary driving force in the formation of a lipid monolayer [35]. With respect to gold, in an early study Smith [62] stressed the effects of carbonaceous contamination on the hydrophilicity of a clean gold substrate turning it hydrophobic. This claim is supported by Gardner and Woods [63] who demonstrated that when organic species are present, the surface of gold is hydrophobic. Therefore, we stress the importance of an adequate cleaning procedure of this sensor substrate to minimize undesirable effects of contaminants present on the working electrode of the QCM crystal. In spite of the fact that avoiding organic and inorganic contamination on the surface of gold is a challenging task, our analysis of the WCA (Table 1) shows that both the Au and $SiO_2$ QCM chips are hydrophilic as a result of our cleaning procedure, where the $SiO_2$ substrate is substantially more hydrophilic. In particular, spreading of vesicles on Au is not straightforward since it has produced conflicting results [35], [41], [64], [65]. Amongst the properties of gold, we find biocompatibility, inert nature, affinity with -thiol group and low electrostatic repulsion to zwitterionic lipids making this noble metal a good candidate for the formation of SLMs, therefore, whether clean Au promotes only vesicle adsorption or adsorption followed by spontaneous rupture was investigated here as a control and was also used as a support for the graphene coatings. Finally, the adsorption of vesicles is influenced by the presence of electrolytes and the ionic strength of the buffer solution. Both parameters have shown to affect the lipid-substrate interaction by means of the charge of the lipids head groups and the net substrate charge [39], [52]. These characteristics altogether constitute the main driving forces to promote vesicle adsorption and eventual fusion of DOPC vesicles to form a uniform membrane [52].

We have investigated the effect of GO and rGO on the assembly of different lipid membrane structures and in combination with the QCM-D technique we proposed them as suitable platforms for the detection of biomolecular interactions. All experiments were performed by following an adapted version of the protocol proposed by Cho and coworkers [36] (for detailed protocol, see ESI). Bare control QCM-D crystals were used in parallel on each of the reported measurements. Results obtained from the control chips were consistent throughout the experimental routines.

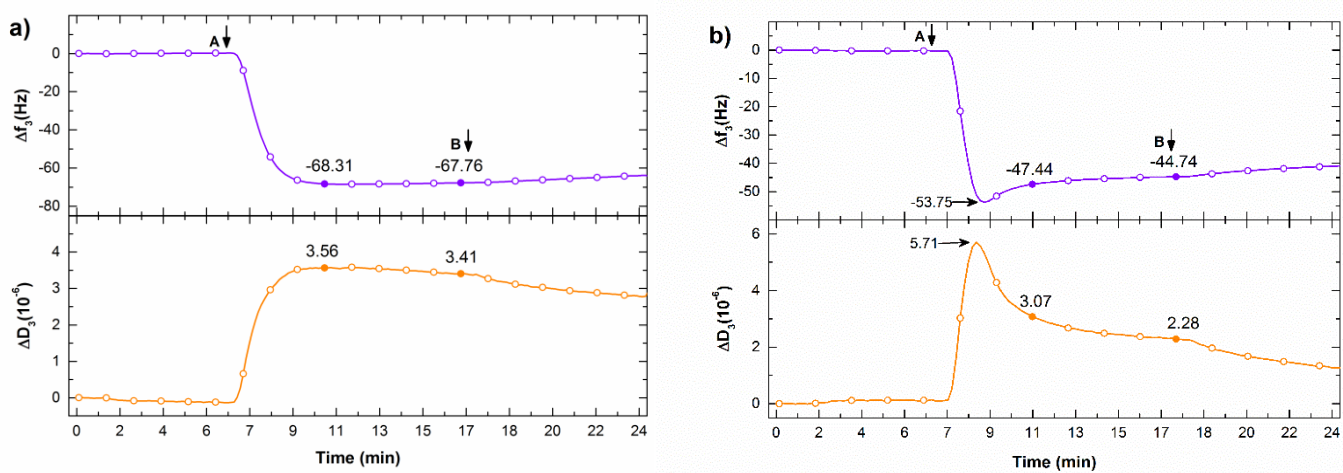

**Fig. 3** Frequency (purple, top) and dissipation (orange, bottom) response from the adsorption of DOPC on **a)** GO-coated SiO$_2$ and **b)** GO-coated Au. Steps: (**A**) DOPC injection and (**B**) buffer rinse.

### 3.2 SLMs on bare substrates
#### 3.2.1 Formation on bare SiO$_2$

First, we report the lipid adsorption kinetics on bare SiO$_2$, as shown in Figure 2(a). After lipid injection (step A), a high frequency downshift indicates that vesicles are adsorbed intact on the surface reaching a critical coverage before they break and spread into a uniform membrane. Within less than 2 minutes the frequency increases stabilizing at -26.65 Hz with a dissipation of 0.99 × 10$^{-6}$ confirming the formation of a bilayer. This energy dissipation value is somewhat higher than the typical reported values during the formation of a SLB (< 0.5 × 10$^{-6}$) [35], [42], [49], [61] denoting that our bilayer is slightly less rigid than previous reports. We emphasize that the observed fast rupture is due to the effect of the Mg$^{2+}$ ions on the reduction of critical coverage thus accelerating the phase of vesicle fusion [39]. The adsorbed lipid remains stable even after a buffer rinse (step B) confirming the formation of a rigid and uniform membrane. After washing out the QCM chip with SDS solution (step C), an increase in the energy dissipation to a median value of 8.52 × 10$^{-6}$ shows that some unabsorbed lipid was removed from the surface. Finally, after rinsing with buffer (step D) the initial baseline is recovered, indicating that the crystal has been fully cleansed. The values for the frequency shift and stabilization throughout the control experiment are in line with the results reported by Keller [35] and Cho [36] on the formation of supported lipid bilayers on clean SiO$_2$ and SiO, respectively.

#### 3.2.2 Formation on bare Au

Similarly, we investigated the adsorption of DOPC vesicles on clean Au. We point at the hydrophilic-hydrophilic interaction between the lipid heads and the substrate as the main force in the formation of a vesicular layer. The lower hydrophilicity from this substrate does not favor any noticeable rupture and a monotonic vesicular adsorption is obtained. The adsorption of intact vesicles on bare Au is corroborated by the report by Liu and Chen [17] in which a supported vesicular layer was required for the evaluation of the rupture of vesicles in the presence of GO. It should be noted that in both reports we used identical Au QCM chips (QSX-301). In addition, the ionic strength from our buffer is not enough to promote vesicle fusion and liposomes adsorbed intact without rupture. This is in contrast with previous reports that have stressed the importance of divalent ions, as Ca$^{2+}$, as mediators in the formation of SLBs on gold [36], [41] i.e. by increasing the deformation of DOPC lipid vesicles [40]. Interestingly, Marques et al. [41] reported the inhibitory effect of NaCl on the formation of lipid bilayers on gold, leading to tubular structures. We indeed tested their experimental conditions (data not shown) using a gold QCM substrate, preparing buffer solution without NaCl and keeping the Mg$^{2+}$ ions. However, the results did not match, perhaps because both the lipid composition (binary lipid vs single lipid) and surface topography differ between our studies. In their study, Marques and coworkers annealed the gold substrate at direct high temperature to obtain smooth micro-domains, while we used the QCM crystals as- received, namely without any other treatment than thorough cleaning, therefore keeping the inherent roughness of the substrate. It has been established that controlling the roughness of a substrate has direct impact on the structure of the membrane to be deposited [66].

The event of intact vesicle adsorption is shown in Figure 2(b) as a high frequency shift (the highest amongst all experiments) of -164.27 Hz and an energy dissipation of 14.08 × 10$^{-6}$, values attributable to the size of vesicles suggesting the high mass loading of the crystal with a non-homogeneous vesicular membrane that releases considerable amounts of energy during its formation and stabilization. In addition, we must consider the added mass of buffer trapped within and between the vesicles. The high dissipation value indicates that the ad layer is not rigid, rather viscoelastic. Finally, after rinsing with buffer (step B) the slight decrease of $\Delta D_3/3$, whilst the frequency remains constant, shows that the vesicles that conform the membrane are close packed and during this stage a lateral shift occurs, as indicated by the energy dissipation variation, with negligible loss or gain of mass. As mentioned before, we highlight the role of the roughness of the Au surface to favor the placement of intact vesicles. AFM topography of our Au QCM crystals is included in the ESI and shows a less smooth surface compared to the SiO$_2$ substrates. In a report from Li et. al. [64] it was established that surface roughness plays a pivotal role in the formation of SLBs on a gold substrate through AFM studies on annealed gold electrodes. During a flame annealing process, clean and large gold grains with atomically flat terraces are produced which help in the promotion of vesicle fusion after lipid deposition. Such atomically smooth surfaces are not present in our crystals. Similar to our results, they indicated having obtained unfused vesicles on rough gold QCM surfaces.

As mentioned before, the electrostatic interaction plays a pivotal role on the formation of SLBs. On this regard, previous reports denoted that negatively charged intact vesicles adsorb onto a titanium oxide or a gold substrate without spontaneous rupture to form SLBs in the absence of divalent ions, like Ca$^{2+}$ [36], [67]. In our study, however, a similar situation occurs with a zwitterionic (neutrally charged) single lipid on bare gold even under the effects of Mg$^{2+}$ ion, which has shown similar strength in the promotion of vesicle rupture [48]. In their protocol, Cho et. al. [36] stressed the necessity for a negatively charged lipid in combination with calcium ions for the formation of SLBs on titanium oxide. As a matter of fact, they stated that their procedure is not suitable for forming single-component zwitterionic SLBs (as DOPC) on titanium oxide, which possesses similar biocompatibility and electrical properties to gold. To overcome this limitation, they proposed and tested the use of an amphipathic AH peptide as a vesicle-destabilizing agent which successfully promoted vesicle deformation and subsequent rupture into a uniform SLB. Overall, SLB formation demands specific surface properties such as surface charge density and hydrophilicity. Therefore, the use of specific agents, including surface modifiers, is a common way to help the promotion of vesicle rupture and GO naturally emerges as a good candidate due to its richness in functional groups that confer this material its distinctive properties. Accordingly, we discuss next the interaction between GO-coated substrates and lipid vesicles.

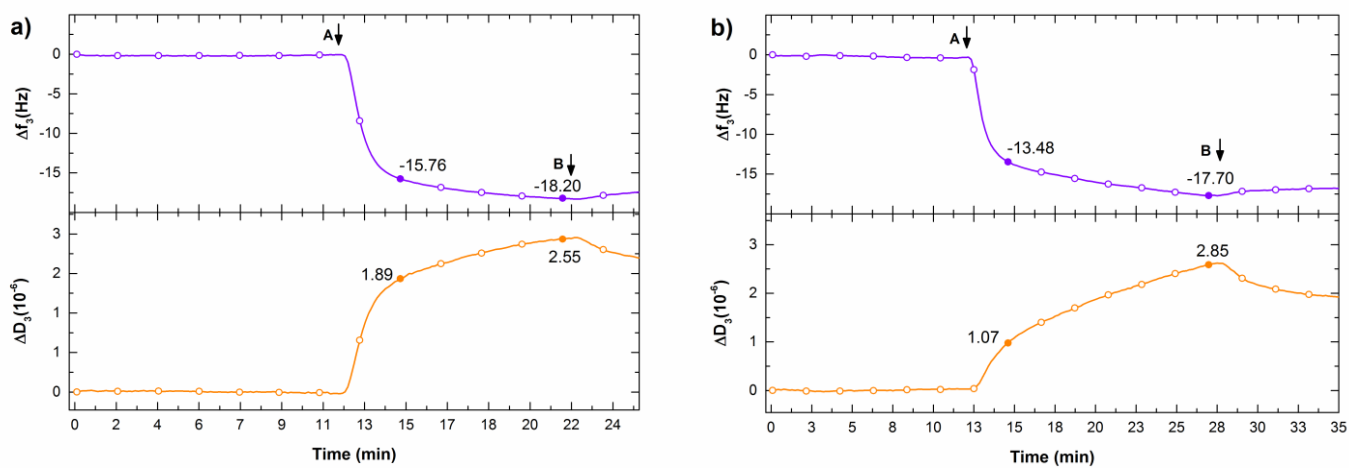

**Fig. 4** Frequency (purple, top) and dissipation (orange, bottom) response from the adsorption of DOPC on **a)** rGO-coated SiO$_2$ and **b)** rGO-coated Au. Steps: (**A**) DOPC injection and (**B**) buffer rinse.

### 3.3 SLMs on GO-coated substrates

According to the widely accepted Lerf-Klinowski model [20], GO sheets have a high number of oxygen groups that confer the material interesting physicochemical properties. Particularly, on the basal plane hydroxyl (-OHs) and epoxy groups give GO its hydrophilic nature [20]. In addition, ionized carboxyl groups (-COOH) at the edges make GO sheets negatively charged [68]. We confirm the presence of these groups in our GO through XPS (see ESI). Previous studies have reported that lipid head groups control the interaction between charged lipids and GO, showing that these have a strong electrostatic interaction with the negatively charged carboxyl groups of GO [28], [69]. Some studies consider that even van der Waals forces might participate on the association between GO and lipids [68]. Specifically, neutrally charged liposomes (as DOPC) associate with the oxidized hydrophilic regions of GO sheets. In addition, system hydration is believed to importantly contribute to the interaction, since water molecules mediate the hydrogen bonding between the carboxyl and the phosphate oxygen present in the lipid headgroup [69]. In a recent work, Willems et. al. [70] have shown, via coarse-grained molecular dynamics simulations, that preformed lipid bilayers and inverted lipid monolayers supported on GO sheets when immersed in water rapidly reorganize into bicelle-like structures. Such rearrangement is thought to be driven by lipid headgroup interactions and was explained by the polarity from the oxygen-containing functional groups in GO. This finding highlights the effect of hydration of the system, thus supporting the vastly accepted hypothesis of the hydrogen bonding between the carboxyl groups of GO and the phosphate head group from DOPC [71]. Willems' and coworkers work preceded a similar study by Rivel et. al. [72] that stressed the competition between the amphiphilic nature of the phospholipid, the hydrophobicity of graphene (also present as hydrophobic domains on GO sheets) and the hydrophilicity of water during the formation of single and multilayer of lipids on graphene.

#### 3.3.1 Formation on GO-SiO$_2$

The formation of lipid membranes supported on GO-coated SiO$_2$ quartz crystals is shown in Figure 3(a). The frequency shift ($\Delta f_3/3$) reaches an initial stabilization value of -68.31 Hz after vesicle injection (step A). Similar to the previous result, we assume the considerable mass uptake due to buffer trapped within the aqueous phase of liposomes and the mass of buffer between them. The energy dissipation value of 3.56 × 10$^{-6}$ indicates the formation of a soft membrane conformed by lipid vesicles which are strongly adhered to the edges of the GO flakes and sparsely distributed on their surface, as it has been previously proposed [69], [73]. From the frequency dissipation response, vesicles adhere without noticeable rupture, as a possible effect of bare SiO$_2$ regions during the coating procedure thus indicating a good coverage of the substrate. In this regard, Furukawa et. al [74] reported that GO blocks the formation of SLBs on a SiO$_2$ substrate where GO flakes are present. This effect is explained by the amphiphilic nature of this graphene derivative due to the presence of both sp$^2$ and sp$^3$ domains. Additionally, it was established by Frost et. al. [45], [75] that liposomes do not rupture when adsorbing to GO. They found that rupture of vesicles is affected by both the dimension of the GO flakes and the diameter of the liposomes. They observed that liposomes fully rupture upon further addition of GO flakes, after they have adsorbed to large GO sheets (0.5-5 µm) obtaining multilayered structures of lipid membranes and GO. Interestingly, their hypothesis is that for vesicle rupture to occur, lipid vesicles must be exposed to two GO sheets, one on each side, where GO sheets are of the same size or larger that the cross-sectional area of the liposome. On this regard, it is evident that in our investigation we exposed only one side of the liposomes to GO flakes when they were adsorbed to the GO-coated substrate hence no rupture would be expected. Our result is in good agreement with a previous report on the adsorption of intact vesicles without fusion on oxidized CVD graphene transferred to a SiO$_2$ substrate [49], however the adsorbed mass in our experiment was higher perhaps due to the higher hydrophilicity of our GO coated crystals, promoting then a higher attraction vesicles. Interestingly, our results are in opposing direction to the findings of Okamoto *et al.* [73] who reported the formation of single and double bilayers via vesicle fusion on GO flakes supported on SiO₂ in the presence of divalent ions, like calcium. We point out at the topographic differences between the substrates used in our studies and their effect on the adsorption and fusion of lipid vesicles. In their study, SLBs were formed by incubation of DOPC vesicles on a GO/SiO₂/Si substrate and analyzed using AFM. During the substrate preparation stage, large blank SiO₂ regions were still present after the deposition of GO solution, while in our study we aimed to obtain a uniform coverage of the substrate. Based on their results, the formation of bilayers and double bilayers on GO sheets is explained as an effect of its surface heterogeneity, however their findings are not conclusive on this matter. As we discussed before, SiO₂ is well known to promote SLB formation due to its hydrophilic property, therefore the strong effect that SiO₂ has on the fusion of DOPC vesicles might be the leading interaction in that scenario. We reason that after the vesicle rupture events originated at the blank SiO₂ regions, fragmented lipid is rapidly attracted by and mobilized onto the regions covered with GO. This hypothesis is supported by previous reports by Hirtz *et. al.* who have shown the self-limiting spreading of DOPC on graphene [34], [76] and, as previously discussed, lipids rapidly reorganize into different lipidic structures in aqueous media [70], [72]. The contrasting homogenous hydrophilicity of SiO₂ and the distributed hydrophobicity from the $sp^2$ domains from GO sheets might lead such interaction. On this regard, the QCM-D technique excels the AFM topographical analysis on the capability to monitor both dynamically and in real-time the adsorption of lipids on specific substrates. It is important to highlight the relevance of vesicular lipid membranes for certain applications, e.g. where lipid vesicles are used as drug carriers and the release of drugs trapped in the aqueous phase must be time controlled through the addition of vesicle destabilizers that promote rupture.

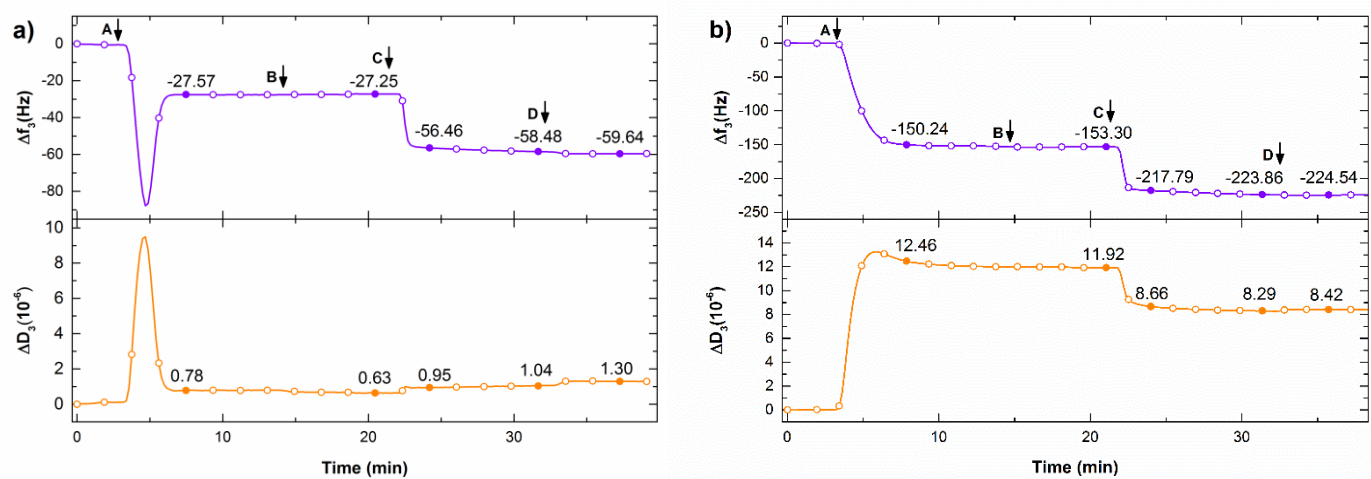

**Fig. 5** Frequency (purple, top) and dissipation (orange, bottom) response from the biotin-avidin binding event **a)** on a lipid bilayer supported on bare $SiO_2$ and **b)** on a vesicular lipid membrane supported on bare Au. Steps: (**A**) DOPC with biotin caps injection, (**B**) buffer rinse, (**C**) avidin injection, (**D**) final buffer rinse.

### 3.3.2 Formation on GO-Au

On the other hand, on the GO-coated gold chip (Figure 3(b)), vesicles are adsorbed intact after lipid injection (step A). The frequency shift reaches a critical coverage point at a value of -53.75 Hz (arrow) indicating the formation of a membrane of intact vesicles which in less than one minute is followed by partial vesicle rupture, shown as a subsequent frequency increase and stabilization to a value of -44.74 Hz before the buffer rinse step (B). After reaching the critical coverage point, the energy dissipation exponentially decreases from a maximum of $5.71 \times 10^{-6}$ to a value of $2.28 \times 10^{-6}$, before rinse. These factors indicate a considerable release of energy during the initial stabilization stage, where vesicles stack then squash and the weakly adsorbed lipid distributed on the surface detaches after buffer rinse (step B), suggesting the gradual compaction of the membrane and a spatial redistribution of the initial vesicular membrane into a different lipidic structure. In contrast to the previous result on GO-coated SiO₂ chip, the lower value of adsorbed mass on the GO-coated Au chip after stabilization indicates that after leaking buffer from within the aqueous phase, some ruptured vesicles reorganize to more likely form bicelle-like structures, as previously discussed from the results on GO in water by Willems and coworkers. Therefore, our results point toward a mixed membrane conformed of intact vesicles and bicelle-like islands.

In both cases, lipid vesicles interact with a heterogeneous surface chemistry present on the GO sheets on these GO-coated chips. We reason that the presence of both hydrophilic/hydrophobic domains establishes an equilibrium on the vesicle-substrate interaction forces which might be altered by the differences on the intrinsic topography of the substrates. Based on our WCAs (Table 1), the GO coating on the smoother SiO₂ substrate is not hydrophilic enough to promote rupture like its bare counterpart while GO on gold, showing slightly more hydrophilic regions than bare gold, leads the adsorption of vesicles followed by partial rupture, as shown in Figure 3(b). In addition, following the previously discussed hypothesis posed by Frost and coworkers, such rupture might occur at specific sites where some vesicles are partially wrapped by GO flakes present on some valleys of the rough gold substrate.

In addition to the hydrophilic vesicle-substrate interaction, we point at the electrostatic force between the negatively charged GO regions and the dipole headgroup of the zwitterionic lipid as the main driving forces for the adsorption of intact vesicles. From our results, it is evident that the ionic strength of the buffer and the cation bonding with the phosphate group of DOPC only promotes partial vesicle rupture upon completion of the critical coverage which in the case of the GO-SiO₂ (Figure 4(a)) the mass loss after rupture is negligible in comparison to the desorbed lipid on the GO-Au chip after partial rupture and stabilization.

### 3.4 SLMs on rGO-coated substrates

Removing the oxygen groups by the thermal reduction of GO coated substrates effectively changes the surface chemistry leading to a highly hydrophobic surface, as seen on the substantial increase on the contact angle after the thermal treatment of the QCM-D sensor set (Table 1). Thus, a new group of measurements were carried out to study the adsorption dynamics of DOPC vesicles on rGO substrates.

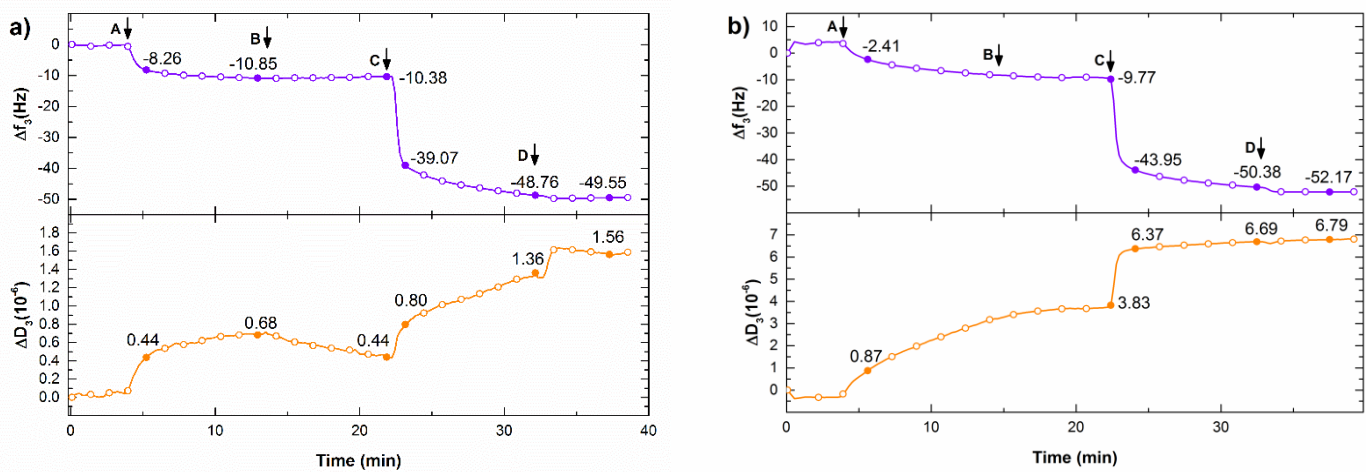

**Fig. 6** Frequency (purple, top) and dissipation (orange, bottom) response from biotin-avidin binding event on a lipid monolayer supported **a)** on rGO-coated SiO$_2$ and **b)** on rGO-coated Au. Steps: (**A**) DOPC with biotin caps injection, (**B**) buffer rinse, (**C**) avidin injection, (**D**) final buffer rinse.

### 3.4.1 Formation on rGO-SiO$_2$

Figure 4(a) shows the monotonic response for the formation of a lipid monolayer on rGO-coated SiO$_2$ chip. After injecting the lipid vesicles (step A), vesicle rupture and spreading into a monolayer occurs. This distinctive adsorption and instantaneous rupture is due to the interaction between the hydrophobic regions of rGO [77] and the hydrophobic fatty acid chains from lipid tails. It has been reported that in the formation of this type of membrane via vesicle fusion on hydrophobic substrates such as a methyl-terminated SAM [35] and CVD graphene [49] a frequency shift of around -13 Hz is expected. In our case $\Delta f_3/3$ varies from an initial stabilization value of -15.76 Hz to a final frequency of -18.20 Hz with a dissipation of $2.55 \times 10^{-6}$ before buffer rinse (step B). These values somewhat differ from previous reports on the formation of uniform lipid monolayers, however they point toward the formation on a non-homogeneous monolayer. On this regard, we hypothesize that individual lipid molecules are attracted during vesicle rupture to the reduced GO sheets with higher hydrophobicity, forming small islands that support an additional lipid membrane on top on them, as it was pictured by Tsuzuki et.al. [27]. Furthermore, the higher mass uptake can be attributed to a wetting film present in the interface. Interfacial water layers have been observed under graphene membranes adhered to sapphire substrate, uniformly trapping water that lifted the edges of graphene sheets, in consequence adding more mass to the sensor [78]. We draw attention to the structural and chemical differences between a transferred CVD graphene sheet and a thermally reduced GO coating. While the first can be considered as a highly crystalline film that might be mono- or few-layer graphene, the second cannot be regarded as a single continuous film, rather a group of overlapping sheets randomly arranged on the surface creating multilayer graphene-like platelets that may preserve different functionalities due to an imperfect thermal reduction process.

### 3.4.2 Formation on rGO-Au

Similarly, on an rGO-coated Au chip (Figure 4(b)), after lipid injection (step A) instantaneous rupture of vesicles occurs with an initial $\Delta f_3/3$ value of -13.48 Hz reaching a final frequency shift of -17.70 Hz indicating the formation of a lipid monolayer membrane. The final dissipation value of $2.85 \times 10^{-6}$ shows that the membrane has viscoelastic properties.

Considering the variations between the initial and final values obtained for the frequency shift in both samples, rGO-SiO$_2$ and rGO-Au, our results are within range of previous reports on the formation of lipid monolayers on hydrophobic graphene [49].

### 3.5 Graphene-SLMs as biomolecular interactions platforms

Finally, we examined the biomolecular interaction between the Biotin-Avidin complex supported by lipid membranes formed on bare and rGO-coated substrates. The biomolecular interaction associated with a vesicular layer, as those from the GO-coated substrates, is out of the scope of this study and has no biotechnical relevance from the perspective of biomimetic membranes.

Our aim was to investigate the kinetics of the lipid adsorption and binding event on the SLMs, especially on graphene due to the similarities between rGO and pristine graphene. On this regard, Hirtz el al. [79] have reported the assembly of inverted phospholipid bilayers (where the hydrophobic tails are facing towards the water/air media and the supporting layer holds the hydrophilic heads) on exfoliated graphene in air, via the dip-pen nanolithography technique. Interestingly, after immersion in buffer of the lipid bilayer they observed a rearrangement into a monolayer with the hydrophilic headgroups facing outwards, more likely happening due to a strong interaction between the hydrophobic surface of rGO and the lipid tails. Obtaining right oriented lipid monolayers is crucial in biomolecular studies in liquid media such as the insertion of peripheral proteins or the present biotin-avidin binding measurement since the interaction can only occur with the biotin molecules attached to the lipid heads.

We have obtained the experimental conditions for real-time monitoring of the detection of biomolecular interactions that can take place in biomimetic membranes supported on graphene through the vesicle fusion technique for the formation of SLMs and employing the QCM-D system. These experiments were performed by following the same process described before and the only difference is the use of 10% biotinylated DOPC lipid vesicles for the formation of the lipid membranes. In addition, we included two extra steps: the injection of avidin protein dispersed in HEPES followed by a final rinse with clean buffer for the elimination of any residual lipid or untied protein.

### 3.5.1 Avidin-biotin binding on bare SiO$_2$ SLM

The first binding event was carried out on a lipid bilayer formed on bare SiO$_2$, shown in Figure 5(a). Initially, the formation of a homogeneous lipid bilayer follows the same adsorption kinetics as described before. After lipid injection (step A) followed by buffer rinse (step B), both values for the frequency and dissipation of -27.25 Hz and $0.63 \times 10^{-6}$, respectively, validate the formation of the bilayer. At step C, avidin injection is followed by an instantaneous binding to the biotin caps attached to the lipid heads. A clean and monotonic mass uptake after protein injection occurs, with a frequency of -56.46 Hz. In addition, a low energy dissipation of $0.95 \times 10^{-6}$ during binding (steps C-D) indicates that the attachment of avidin molecules happens without altering the structure of the previously formed rigid bilayer. Here the difference between the stable frequency value before protein injection and the final frequency at the binding event completion (step D) is $\approx 29\ Hz$ for the bilayer support.

**Table 2** Summary of results

| Crystal | Coating | $\Delta f_3/3$ [Hz] | $\Delta D_3/3$ [$\times 10^{-6}$] | Stabilisation time | Structure type |
|---|---|---|---|---|---|
| SiO$_2$ | Bare | -26.73 | 1.00 | 10:15 | Lipid bilayer |
|  | GO | -67.76 | 3.41 | 10:00 | Intact vesicles layer |
|  | rGO | -18.20 | 2.55 | 9:53 | Monolayer |
| Au | Bare | -169.96 | 13.31 | 10:12 | Intact vesicles layer |
|  | GO | -53.75 → -44.74 | 5.71 → 2.28 | 10:24 | Intact vesicles + Bicelle-like structures |
|  | rGO | -17.70 | 2.85 | 14:58 | Monolayer |

### 3.5.2 Avidin-biotin binding on bare Au SLM

In contrast, the binding event supported on a bare Au chip (Figure 5(b)) shows a frequency a value of -153.30 Hz at the end of the stabilization region (steps B-C) after the adsorption of intact vesicles (step A). Then, the frequency reaches a value of -217.79 Hz after avidin injection (step C) with a final frequency value of -223.86 Hz before rinse. This variation corresponds to $\approx 70\ Hz$. We validate this high value from the surface area of the vesicular membrane, where a myriad of biotin molecules is present at the outer layer of each liposome for the protein to bind. The overtones considerably diverged (see ESI) throughout the experiment indicating that there is frequency dependence during the shear mode of the crystal. A high dissipation value of $12.46 \times 10^{-6}$ is reached after a slight peak during the formation of the supported lipid vesicles while $\Delta f_3/3$ varies monotonically. This behavior is indicative of vesicle-vesicle compaction prior to the formation of a non-homogeneous soft membrane. In contrast to the vesicular lipid membrane formed on bare Au (Figure 2(b)) the frequency shift during stabilization is lower and no partial rupture occurs, perhaps due to the hydrophobic nature of biotin [55] present in the lipid heads, changing the vesicle-substrate interaction by the reduction of the hydrophilic attracting force, and thus reducing the number of adsorbed vesicles. Finally, after rinse (step D), the slight increase in both $\Delta f_3/3$ and $\Delta D_3/3$ to values of -224.54 Hz and $8.42 \times 10^{-6}$, respectively, indicates that some remaining unattached protein is dragged from saturated sites to available biotin sites to attach, similar to the previous result.

### 3.5.3 Avidin-biotin binding on rGO-SiO$_2$ SLM

Next, the lipid monolayer formed on rGO-SiO$_2$ was used as a supporting platform for the protein binding (Figure 6(a)). The frequency dissipation before avidin injection (step A) stabilized at -10.38 Hz. We consider this value within the lower limit for the formation of a lipid monolayer on graphene, according to previous reports [49]. The low energy dissipation value of $0.44 \times 10^{-6}$ with a momentary increase to $0.68 \times 10^{-6}$ during the monolayer formation (steps A-C), indicates that the membrane is being compacted to be finally closely attached to the surface. The noticeable unsteady value of $\Delta D_3/3$ throughout the lifetime of the experiment (~38 min) suggests that the membrane is settling onto the surface. However, the uniformity of $\Delta f_3/3$ throughout the stabilization period (steps B-C) indicates that the membrane is rigid and the movement is parallel to the shear force of the QCM-D sensor. After injection of avidin (step C) the instantaneous frequency shift indicates a successful binding event reaching an initial value of -39.07 Hz then stabilizing at -49.55 Hz after buffer rinse (step D). These values represent a variation of $\approx 40\ Hz$.

### 3.5.4 Avidin-biotin binding on rGO-Au SLM

Finally, Figure 6(b) shows the adsorption kinetics of the binding event on a lipid monolayer supported on a rGO-Au chip. At the end of the membrane formation stage (steps A-C) the frequency stabilizes to a value of -9.77 Hz which is close to the lower limit of the formation of a monolayer. We reason that this low value of $\Delta f_3/3$ indicates an incomplete coverage of the substrate and the formation of rGO patches supporting islands of lipid monolayer. In this case, the frequency variation is of $\approx 42\ Hz$, which is slightly higher than the values obtained for bilayer and monolayer on SiO$_2$. This difference can be explained from a structural point of view of the distribution of the monolayer patches on the substrate. We consider that they are spatially separated by the topography of the Au substrate, leaving exposed biotin caps attached to the lipid heads that are located at the margins of the lipid membrane, therefore increasing the available binding sites, whereas for the monolayer and bilayer on SiO$_2$ the avidin binds to the biotin present in the superficial layer of the continuous membranes. In addition, the increase in the dissipation after the binding event (step C) and buffer rinse (step D) to a final value of $6.79 \times 10^{-6}$ shows that the membrane is not homogeneous. This behavior might corroborate our hypothesis for the structural distribution of lipid patches since the buffer flow hits them generating lateral movement, therefore increasing the energy dissipation.

## Conclusions

We have described the construction and characterization of graphene supported biomimetic lipid membranes and the biomolecular interactions that can take place on different substrates. We elucidated that both the chemical heterogeneity of GO and the nature of supporting substrate lead to different lipid structures obtaining either a membrane of intact vesicles or a mixed layer composed of intact vesicles and ruptured vesicles that reorganised into structures that resemble lipidic bicelles. In contrast, lipid monolayers were successfully obtained on both substrates coated with reduced GO, where the hydrophobicity of the supporting material reached its highest point. In general, the formation of the range of lipidic structures presented on this work occurred in less than 15 minutes in terms of the time to become stable or for the time the response showed a monotonic behaviour.

From a general standpoint, the variability of the responses obtained for the interaction between lipids and coated substrates in the four cases under study that involved graphene is appreciable as presented in Table 2, i.e., GO-SiO2, GO-Au,

rGO-SiO2 and rGO-Au. This contrast of responses is explained not only from the perspective of the lipid-graphene interface but also from the effect that the underlying layer has on the graphene support. The wetting transparency effect [80] has been demonstrated in terms of the permeability of the graphene film to the hydrophilicity/hydrophobicity of the support substrate and the role that the interfacial water layer plays, transferring these forces from the substrate to the graphene surface [78]. This fact may explain with enough consistency the differences on the formation of SLMs using the same support substrate and attributable to the difference in surface chemistries.

Graphene stands as an effective route for the chemical modification of the selected substrates and as the underlying platform for the development of a mass sensitive biosensor. The interaction between lipids and graphene is strongly led by electrostatic and hydrophobic interactions between them, therefore is necessary to investigate the crucial experimental parameters to obtain reproducible and defect free membranes. The techniques for the formation of lipid monolayers have evolved, moving from the Langmuir-Blodgett technique [81] towards the vesicle fusion [35], [50] on substrates with inherent or modified hydrophobicity. Performing the latter on graphene sheets is a straightforward option for the site-selective formation of lipid monolayers due to the strong interaction between the lipid tails and the graphene plane.

Overall, our results shed light on the interaction between zwitterionic lipids and the dynamics of the physisorption to graphene platforms for potential biotechnical applications. Despite lipid monolayers do not resemble the complexity of biological membranes due to their structural simplicity, they have shown great utility on the evaluation of the interfacial organization of lipid membrane constituents and the changes in the interfacial organization upon the insertion of amphipathic compounds [82] and due to their homogeneity, stability and planar geometry lipid monolayers have been proposed over bilayers as suitable models to characterize protein-membrane interactions [83].

Following the monolayer technique, lipid monolayers have been used for the incorporation of amino acids, like antimicrobial peptides, or proteins, like cardiotoxins as their site of actuation is at the cell membrane level, binding and disrupting the outer membrane [83]. This specific affinity to monolayers could potentially increase the sensitivity of the binding detection in comparison to a bilayer membrane.

Is of our interest to use the proposed rGO-QCM-D platforms as a biomimetic device to study the insertion and binding mechanism of tail-anchored proteins and Odorant-Binding Proteins (OBPs), a soluble protein secreted in the nasal mucus of animal species and in the sensillar lymph of chemosensory sensilla of insects. Our results will serve as the basis to achieve such biosensing system. In the future, our work will be undertaken to study the viscoelastic properties of the adsorbed membranes and binding events using models such as the Sauerbrey equation and the Voigt model to characterize the thickness and mass of the ad layers.

## Acknowledgements

DM acknowledges The National Council for Science and Technology (CONACyT), Mexico for the financial support. AV, AFV and SG acknowledge funding from the Engineering and Physical Sciences Research Council (EPSRC) grants EP/K016946/1 and EP/G03737X/1. The authors acknowledge E. W. Hill and B. Grieve for helpful discussions.

# Adsorption and binding dynamics of graphene-supported phospholipid membranes using the QCM-D technique

D. A. Melendrez, T. Jowitt, M. Iliut, A.F. Verre, S. Goodwin and A. Vijayaraghavan

**Supplementary Information**

**Experimental**

**Overview**

The successful formation of uniform supported lipid membranes demands following a standardized procedure. Here, we describe the experimental steps to prepare Small Unilamellar Vesicles (SUVs), to condition the QCM-D system, present them to selected substrates and acquire the frequency and dissipation responses for further analysis.

**Reagents**

For Graphene Oxide (GO) preparation: Hydrogen Peroxide ($H_2O_2$) 30% (Sigma Aldrich), Sulphuric acid ($H_2SO_4$) 98% (Sigma Aldrich), Sodium Nitrate ($NaNO_3$) 98% (Alfa Aesar), Potassium Permanganate ($KMnO4$) 98% (Alfa Aesar)

For buffer preparation: Milli-Q water (>18 M$\Omega$), 1,2-dioleoyl-sn- glycero-3-phosphocholine (DOPC) lipid (Avanti Polar Lipids), Biotinyl Cap dispersed in Chloroform (10 mg/mL, Avanti Polar Lipids), Avidin protein from egg white (Sigma Aldrich), Analytical grade HEPES buffer (Acros Organics), NaCl (powder, Sigma Aldrich), $MgCl_2$ (powder, Sigma Aldrich), NaOH (pellets, Fischer Scientific), $H_2O_2$ (30% solution), Ammonia (25% solution, Sigma Aldrich),

For cleaning: Sodium dodecyl sulphate (Fischer Scientific), Hellmanex II (Hellma Analytics).

**Graphene oxide preparation**

Graphene oxide used in this work was prepared according to a modified Hummers method described in ref. (1). Briefly, graphite flakes of 50 mesh (1g) and $NaNO_3$ (0.9 g) were mixed in concentrated $H_2SO_4$ (34 ml) in a round bottom flask and kept overnight to intercalate. Then the mixture was cooled down in an ice bath and 4.5 g $KMnO_4$ where added slowly under constant stirring. The resulting mixture was left for 5 days at RT for graphite oxidation. After oxidation process was complete, the resulting brown slurry was diluted at a slow rate with 100 ml $H_2SO_4$ solution of 5% after which 10 ml of $H_2O_2$ solution of 30% was added dropwise. Finally, the dispersion was further diluted with 100 ml mixture of $H_2SO_4/H_2O_2$ of 3%/0.5%. The resulted graphite oxide was purified via centrifugation process by repeated washing with diluted $H_2SO_4$ and then DI water until the pH of the supernatant was close to neutral. The homogenisation and complete exfoliation of graphene oxide was performed using a vertical stirrer at a low speed for ~1h. The stock solution of GO (8.1 mg/mL) was diluted to a concentration value of 0.5 mg/mL.

**Buffer solution**

The buffer solution is prepared diluting 10 mM HEPES, 100 mM NaCl, and 5mM $MgCl_2$ in MilliQ water. The pH is adjusted to 7.4 with a 1M NaOH solution when necessary. Stir this solution for at least 2 hours to ensure complete dissolution. To increase the pH, add dropwise the sodium hydroxide solution during gentle stirring until a stable value is reached. Filter the buffer with the 0.2 µm nylon membranes. Store the buffer in the fridge for up to two weeks.

**Cleaning solutions**

For cleaning the QCM-D system and quartz crystals, prepare both 2% SDS and Hellmanex II solution in MilliQ water. A strong cleaning solution for gold crystals (QSX-301) is prepared as a 5:1:1 mixture of MilliQ water, Ammonia (25%) and Hydrogen Peroxide.

**Lipid vesicle preparation (and Biotin caps incorporation)**

To obtain DOPC SUVs follow the next procedure.

Thoroughly rinse the inner walls of a 5-mL glass vial with chloroform using chloroform syringes. Dry the vial using a soft beam of $N_2$. Take 1 mL from DOPC lipid dispersed in chloroform (2.5 mg/mL) and pour it in the clean vial. *Note*: for Biotin caps incorporation, take 25 µL of this vitamin dispersed in chloroform and mix it in the same vial. Dry the chloroform with a soft beam of $N_2$ until complete evaporation. Hydrate the lipid (/vitamin) with 1 mL of HEPES buffer solution. Fill a 1 mL extruder syringe with the hydrated lipid (/vitamin). Place one 50 nm polycarbonate filtering membrane (Nalgene) at the middle of the Teflon receptacle of the extruder, add two spacers per side and tightly close the hex nut. Insert another clean and empty 1 mL syringe on the opposite side of the Teflon receptacle. Extrude the dispersed lipid for at least 23 times. Be gentle to avoid tearing the filtering membrane. It is recommended to use freshly made lipid vesicles to avoid vesicle aggregation.

**Initial preparation of Quartz Crystals**

These cleaning procedures are based on the protocol provided by QSense (2). The sensors used in this study are QCM with gold surface (QSX-301) and with Silicon Dioxide (QSX-303)/Silicon Dioxide 300nm (QSX-318). A teflon QCM cleaning holder (Q-Sense, QCLH 301) was used to prevent scratches on the surface.

**$SiO_2$ QCM chips**

The following cleaning steps are regarded as mild cleaning and also applies for the gold crystals as a routine cleaning procedure.

1. UV/ozone treat for 10 minutes.
2. Immerse the sensor surfaces in the solution of 2% SDS and sonicate them for 15 minutes.
3. Rinse the sensors with abundant MilliQ water.
4. Immerse the sensors in MilliQ water and sonicate them for 15 minutes.
5. Immerse the sensors in 99% ethanol and soak them for 10 minutes.
6. Dry the surfaces using a mild beam of nitrogen gas.
7. UV/ozone treat for 25 minutes.

**Au QCM chips**

- Chemical treatment (Ammonium Peroxide Mix)

This cleaning process should be carried out under a fumes hood, wearing adequate PPE.

1. Using wash bottles, squirt the following solutions over the working electrode: 10% Decon 90, acetone and isopropanol. Use DI water between each solution to rinse well the surface.
2. Treat the crystals under UV/ozone atmosphere for 10 minutes.
3. Heat the strong cleaning solution for gold to 75 º C.
4. Place the sensor in the heated solution for 5 minutes.
5. Rinse the sensors with MilliQ water. Keep the surfaces wet after ammonium-peroxide immersion until they are rinsed well with water.
6. Dry with nitrogen gas.
7. UV/ozone treat for 25 minutes.

- Surfactant treatment

Follow the same cleaning steps for silicon dioxide sensors.

**Coating QCM-D crystals with GO**

In order to coat the Au/$SiO_2$ crystals, the following methodology must be applied after completing the appropriate cleaning procedure.

1. Configure the following parameters in the *spin coating machine (SCM)* (Laurell technologies Corp. WS-650MZ-23NPPB)

    Speed = 3500 rpm, acceleration = 350 rpm/sec, time = 120 sec.

2. Place a QCM crystal in the vacuum nuzzle of the SCM.
3. Drop cast 70 μl of GO (0.5 mg/ml) on the surface of the Au/$SiO_2$ QCM working electrode. Let the solution settle for 30 seconds.
4. Close the lid of the SCM and start the spinning.
5. Repeat steps 2-4 once for each set of crystals.

**Thermal reduction of GO-coated crystals**

This procedure requires to preheat the oven before cleaning and coating the desired number of sensors since the heating curve from each oven may vary. In our case, the oven was set at 180ºC to preheat 1 hour before placing the chips to be reduced.

Follow the next procedure immediately after the spin coating of crystals is completed.

1. Distribute the selected chips to be reduced in petri dishes and label them accordingly.
2. When the over reaches 180 ºC place the petri dishes with the chips inside the chamber and close the door tightening the screw to ensure good vacuum.
3. Leave the samples for 20 hours.
4. After 20 hours shut down the heater, turn off the vacuum pump and slowly turn the intake valve from the vacuum oven to the *open* position. *Note*: to avoid blowing away the chips inside the oven, turn the valve gently until the atmospheric pressure fills the chamber.
5. Let the chamber cool down for a few minutes and wearing heat gloves carefully take the QCM chips with Teflon tweezers holding them from the edges.
6. Store the crystals in order inside holding boxes for a safe transport.

**QCM-D measurement procedure**

**Initial system cleaning and priming**

The following steps are intended to be applied on the Q-Sense Omega Auto (Biolin Scientific) system, which consists of 8 sensing ports automatically fed through customized scripts.

- Thorough ports and tubing cleaning

a. Load all ports with clean maintenance sensors. *Note*: verify the right position of the sensor matching the anchor symbol.
b. All the ports (1-8) must be initially washed by running 2% SDS at a flow rate[1] of 25 $\mu$L/min for at least 10 min. This step should remove all remaining lipids and biological material from all tubing and syringes.
c. Rinse with system liquid[2] for at least 15 min.
d. Flow 2% Hellmanex through the system for at least 10 min.
e. Finally rinse with system liquid for at least 15 min.
f. Remove the maintenance sensors, rinse the chamber with MilliQ water.

- Sensors & ports priming

Eliminating trapped bubbles is crucial to obtain a stable baseline and a steady response via continuous buffer flow.

a. Load the chamber with the desired number of sensors to be used. Up to four chips can be used for a parallel data acquisition. Ensure that the electrodes are all dry during placement to avoid any variations during measurements.
b. Set the chamber temperature to 24 ºC to minimize thermal drift.
c. Run system liquid through the loaded ports until a stable baseline is noticeable. *Note*: despite this step can be programmed to automatically stop when a stable baseline is reached it is recommended to run it manually to override the baseline criteria from the system.
d. Vacuum ports and start running buffer solution through the working sensors for at least 5 min before the actual sample injection and *vesicle fusion* technique is applied.

---

[1] All flow rates are equal to 25 $\mu$L/min unless specified.
[2] *System liquid* refers to ultrapure water.

**Formation of supported lipid membranes using the vesicle fusion technique**

The aim of these steps is to present lipid vesicles to QCM-D sensors with different working-electrode surfaces by flowing Small Unilamellar Vesicles (SUVs) dispersed in buffer solution. An initial vesicle-substrate interaction is expected to be followed by a vesicle-vesicle interaction to obtain specific structures of lipid membranes. This procedure has been successfully applied to obtain bilayers on clean hydrophilic substrates such as SiO$_2$ and monolayer on modified Au (3). The adsorption and formation of all the lipid membranes discussed in this work was accomplished by following the same experimental steps.

1. Deposit the extruded lipid vesicles dispersed in buffer in a 1.5 mL vial and place it in the right-hand rack and lock the lid from the Omega Auto system. Note: in case of carrying out a binding event, also place in the rack a vial containing 100 µL of protein dispersed in chloroform, then dried and finally hydrated with 900 µL of HEPES buffer.

2. Run buffer solution through the desired chips for at least 5 min. Start recording the frequency and dissipation responses from this step. Note: verify the stability and flatness of the baseline during this time. In case that some harmonics show jumps and high variation this may indicate the presence of bubbles in the system and/or a bad interface between the working electrode and the media. To solve this, redo steps c and d from the priming section.

3. After 5 min of stability inject the lipid vesicles (0.1 mg/mL) to the desired chips during 10 minutes. Note: The system will indicate the quantity of lipid in buffer solution required to complete this step, however 1.0 mL should be enough to run 4 parallel measurements with the same parameters described here. In case that the vial is not filled to the right level, the system will automatically stop the execution of the script.

4. *Verify* the resonant frequency and dissipation values in liquid. Right after lipid injection a frequency shift must occur showing some mass uptake and an increase in the energy dissipation. Depending on the type of membrane being formed the frequency shift will stabilize to a specific value. *Note*: using a control chip is highly recommended to verify the validity of the experimental procedure. A well-known adsorption kinetics is that for a bilayer on bare SiO$_2$ where the values from Fig. 2a (main text) are expected, with a tolerance of $\Delta f \pm 1$ Hz and $\Delta D \pm 0.5 \times 10^{-6}$.

5. Rinse the lipid layer with buffer for at least 5 min to remove any excess lipid and homogenize the membrane.

6. If performing the binding measurement, inject the Avidin from the vial (after SLM stabilization) and let it settle under continuous flow for at least 5 min.

7. Rinse all sensors with buffer to eliminate any excess protein and/or material deposited on the surface and to record complete values for further analysis.

8. When the main body of the analytical script is completed, the system will run a wash routine. During these steps, all the material present on the sensors will be removed and the surfaces, syringes and tubing will be washed using the selected surfactants (Hellmanex and/or SDS) and finally rinsed with system liquid.

9. Upon completion, the door can be opened and the sensors can be taken out of the chamber.

10. It is recommended to leave the chamber clean and dry to be ready to use in subsequent experiments.

**Samples characterization**

The atomic force microscopy (AFM) of the GO and rGO was performed using a Bruker Dimension FastScan probe microscope operating in taping mode. The tips used for the surface scanning were aluminium coated silicon FastScan-A tips from Bruker. For coated crystals characterization, the diluted GO dispersion (0.5 mg/mL) was casted on clean QCM-D substrates and spin coated as previously described.

The scanning electron microscopy (SEM) was performed on a SEM Zeiss Ultra setup, using an accelerating voltage of 5 kV.

The X-ray photoelectron spectroscopy (XPS) data were collected on a SPECS custom built system composed of a Phobios 150 hemispherical electron analyser with 1D detector. The X-ray source is a microfocus monochromated Al K-alpha (1486.6eV) source. All spectra were collected with a pass energy of 20eV. Combined ultimate resolution as measured from Ag 3d is 0.5eV with X-ray source and 20eV pass. The XPS data processing was done using CasaXPS software (version 2.3.16 PR 1.6). The C1s region peak fitting was done using Gaussian/Lorentzian shape components (for sp$^3$ carbon) and asymmetric shape components (for sp$^2$ carbon) respectively. XPS C1s region was fitted with the synthetic components in the manner which minimizes the total square error fit and corresponds to the literature reports. In the case of rGO, it was impossible to distinguish between sp$^2$ and sp$^3$ carbons, therefore the signal was fitted with a single asymmetric component. The GO sample for XPS was prepared by drop casting the dispersion on a clean Si/SiO2 (300nm) and drying in a vacuum oven to achieve a film thickness not less than 10 nm. The rGO sample was prepared using the same conditions used for the reduction of GO on QCM crystals. The GO vas first casted and dried on the Si/SiO2 (290 nm) substrate, followed by the reduction in vacuum at 180 ºC for 20 hours.

Raman spectrum was taken on a Renishaw Raman system equipped with a Leica microscope and a CCD detector. Raman spectrum was recorded using 532 nm laser line (Cobolt SambaTM continuous wave diode-pumped solid-state laser, 20 mW), and the laser power was kept below 10 µW to avoid thermal degradation of the samples. 30 spectra per sample was taken. The relative intensity ratio ($I_D/I_G$) was measured from the averaged acquired mappings.

**Results and discussion**

**Contact angle**

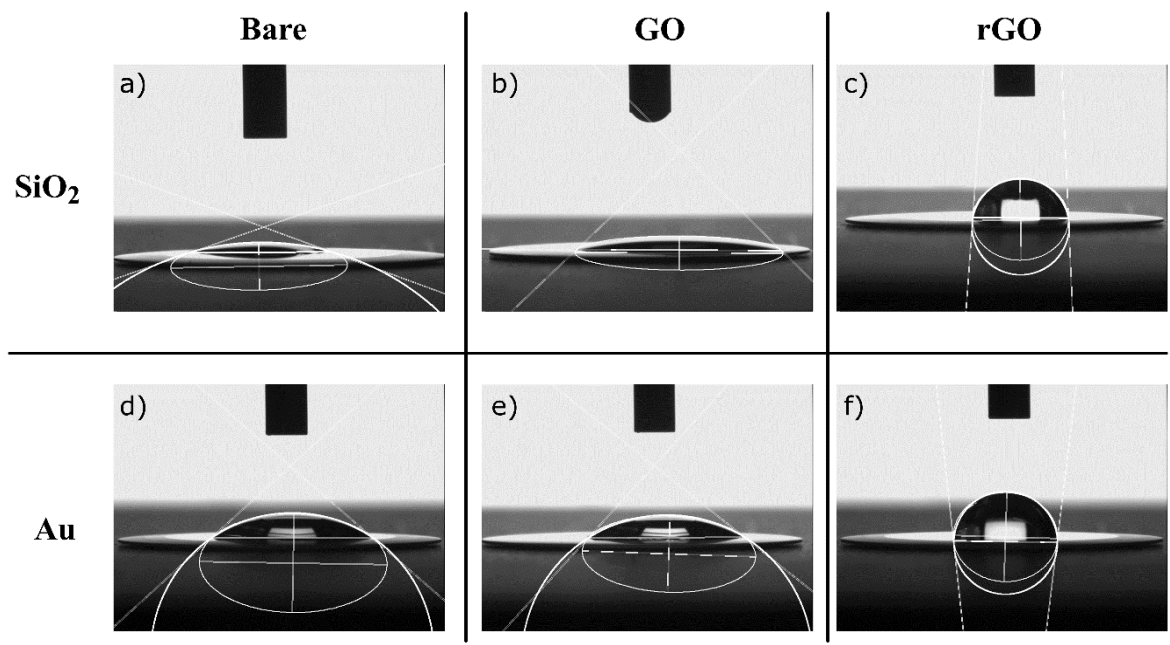

**Fig.S1** Contact angle sheet. **a-f**) Manual fit of the water droplet using the ImageJ plugin [ref]

The wetting contact angles for the range of QCM crystals is shown in Fig. S1. The manual circle-ellipse fittings were computed using an ImageJ software plugin developed and published by Marco Brugnara (4) for such specific task. The software works on pre-captured high contrast images of sessile drops which are processed by first inverting the image upside down, namely, the water droplet must be pending from the top of the image, then two points are selected for the baseline of the droplet and finally three edge points that follow the curvature of the droplet are selected. On each case, 5 readings were captured for statistical effects and the results given by the script are shown in Tables S1 and S2. Table S1 shows the results for the $SiO_2$ crystal variations (Fig. a-c) while Table S2 shows the results obtained for the Au crystal variations (Fig. d-f). In both cases the highlighted cells show the final average value for the ellipse fitting from which the standard deviation showed a lower value than that obtained for the circle fitting.

| Crystal type | Theta C | Uncertainty | Theta Left | Theta Right | Theta E | Circle StDev | Ellipse StDev |
|---|---|---|---|---|---|---|---|
| SiO₂ - bare | 15.8 | 0.1 | 20.7 | 18.8 | 19.8 | 8.56E-02 | 8.92E-04 |
| | 14 | 0.1 | 16.6 | 13.5 | 15 | 1.16E-01 | 2.98E-03 |
| | 14 | 0.2 | 18.6 | 14 | 16.3 | 2.03E-01 | 2.27E-03 |
| | 13.3 | 0.1 | 14.9 | 13.4 | 14.2 | 1.06E-01 | 2.73E-03 |
| | 15.1 | 0.2 | 21.8 | 21 | 21.4 | 2.35E-01 | 2.15E-03 |
| Averages | **14.44** | **0.14** | **18.52** | **16.14** | **17.34** | **1.49E-01** | **2.20E-03** |

| Crystal type | Theta C | Uncertainty | Theta Left | Theta Right | Theta E | Circle StDev | Ellipse StDev |
|---|---|---|---|---|---|---|---|
| SiO₂ - GO | 25.3 | 0.3 | 38.3 | 34.4 | 36.4 | 4.31E-01 | 4.19E-04 |
| | 24.8 | 0.3 | 35.4 | 34.7 | 35 | 5.50E-01 | 6.91E-04 |
| | 24.2 | 0.2 | 30.8 | 33 | 31.9 | 3.40E-01 | 9.99E-04 |
| | 24.9 | 0.1 | 30 | 27.7 | 28.8 | 2.38E-01 | 7.16E-04 |
| | 25 | 0.2 | 25.3 | 31.7 | 28.5 | 3.72E-01 | 6.09E-04 |
| Averages | **24.84** | **0.22** | **31.96** | **32.3** | **32.12** | **3.86E-01** | **6.87E-04** |

| Crystal type | Theta C | Uncertainty | Theta Left | Theta Right | Theta E | Circle StDev | Ellipse StDev |
|---|---|---|---|---|---|---|---|
| SiO₂-rGO | 82.6 | 0.6 | 89.2 | 92.6 | 90.9 | 8.19E-01 | 1.16E-04 |
| | 82.8 | 0.4 | 88.5 | 89.2 | 88.8 | 5.10E-01 | 5.43E-04 |
| | 78.8 | 0.4 | 86.3 | 89.2 | 87.8 | 4.90E-01 | 3.60E-04 |
| | 79.5 | 0.3 | 87.2 | 86.2 | 86.7 | 3.64E-01 | 5.18E-04 |
| | 80.8 | 0.5 | 87.9 | 88 | 88 | 6.14E-01 | 5.47E-04 |
| Averages | **80.9** | **0.44** | **87.82** | **89.04** | **88.44** | **5.59E-01** | **4.17E-04** |

**Table S1** Manual fitting results for SiO₂ crystal set using the Contact Angle ImageJ plugin. Shadowed cell value is the final angle.

| Crystal type | Theta C | Uncertainty | Theta Left | Theta Right | Theta E | Circle StDev | Ellipse StDev |
|---|---|---|---|---|---|---|---|
| Au - bare | 34 | 0.2 | 42.2 | 42.2 | 42.2 | 3.10E-01 | 3.86E-04 |
| | 34.4 | 0.3 | 41.6 | 39.7 | 40.7 | 3.82E-01 | 4.88E-04 |
| | 34.3 | 0.3 | 38.7 | 41.2 | 40 | 3.99E-01 | 8.96E-04 |
| | 34.2 | 0.3 | 40.9 | 42.5 | 41.7 | 4.51E-01 | 7.15E-04 |
| | 34 | 0.3 | 40.3 | 43.6 | 42 | 4.62E-01 | 4.44E-04 |
| Averages | **34.18** | **0.28** | **40.74** | **41.84** | **41.32** | **4.01E-01** | **5.86E-04** |

| Crystal type | Theta C | Uncertainty | Theta Left | Theta Right | Theta E | Circle StDev | Ellipse StDev |
|---|---|---|---|---|---|---|---|
| Au - GO | 32.2 | 0.3 | 38.8 | 40 | 39.4 | 3.88E-01 | 5.83E-04 |
| | 32.7 | 0.2 | 40.9 | 39.4 | 40.2 | 3.59E-01 | 7.46E-04 |
| | 32.3 | 0.2 | 36.8 | 38.4 | 37.6 | 2.61E-01 | 8.18E-04 |
| | 32.2 | 0.2 | 35 | 38.1 | 36.6 | 3.26E-01 | 8.88E-04 |
| | 32 | 0.3 | 35.6 | 37.9 | 36.8 | 4.86E-01 | 4.74E-04 |
| Averages | **32.28** | **0.24** | **37.42** | **38.76** | **38.12** | **3.64E-01** | **7.02E-04** |

| File Name | Theta C | Uncertainty | Theta Left | Theta Right | Theta E | Circle StDev | Ellipse StDev |
|---|---|---|---|---|---|---|---|
| Au - rGO | 87.4 | 0.3 | 94.6 | 93.9 | 94.2 | 4.74E-01 | 5.58E-04 |
| | 87.2 | 0.4 | 97.1 | 96.1 | 96.6 | 6.32E-01 | 1.18E-04 |
| | 87.2 | 0.4 | 95.7 | 94.4 | 95.1 | 5.23E-01 | 8.16E-05 |
| | 88.1 | 0.3 | 94.9 | 95.8 | 95.4 | 5.06E-01 | 2.56E-04 |
| | 86.7 | 0.3 | 92.2 | 92.6 | 92.4 | 4.11E-01 | 1.69E-04 |
| Averages | **87.32** | **0.34** | **94.9** | **94.56** | **94.74** | **5.09E-01** | **2.37E-04** |

**Table S2** Manual fitting results for Au crystal set using the Contact Angle ImageJ plugin. Shadowed cell value is the final angle.

**SEM images**

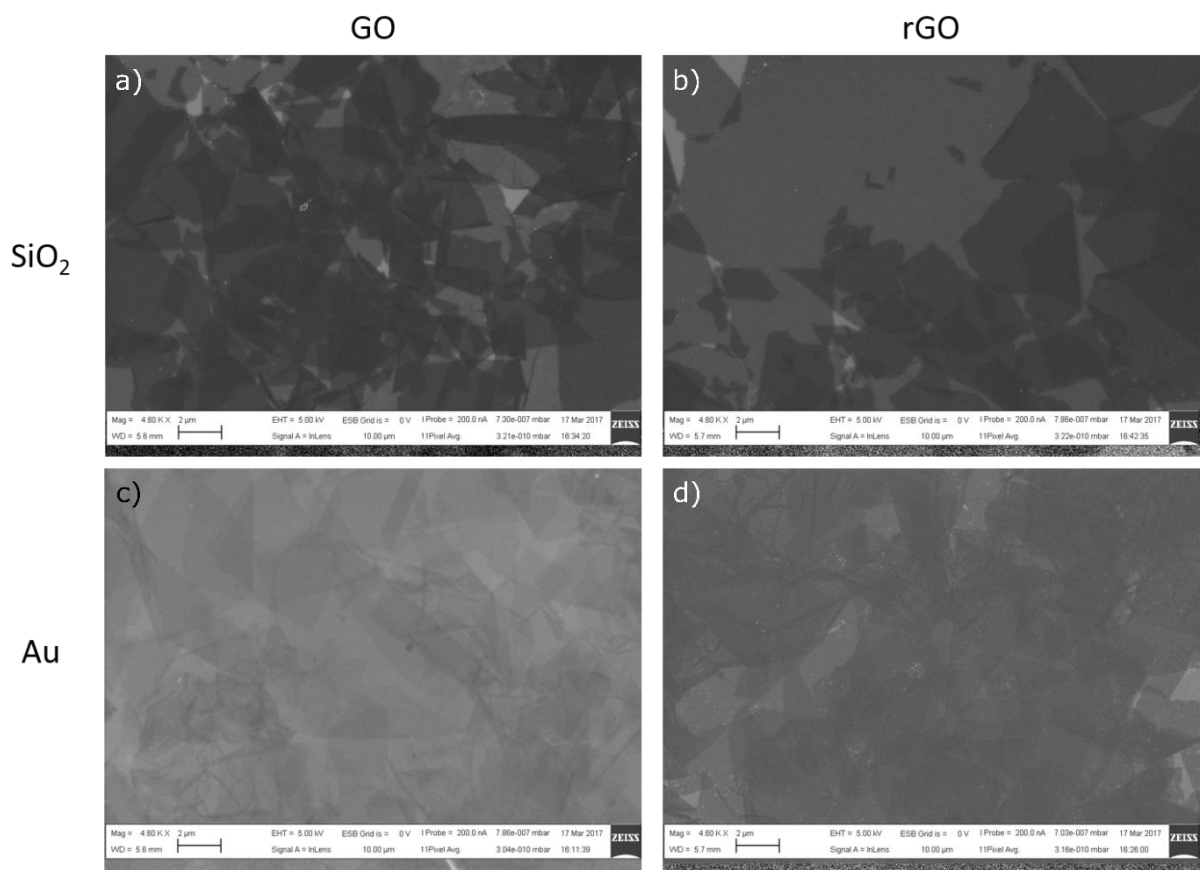

**Fig.S2** Scanning Electron Microscope (SEM) showing GO and rGO flakes arrangement on: **a)** on GO-SiO$_2$ **b)** rGO-SiO$_2$, **c)** GO-Au **d)** rGO-Au. Probe voltage 5.00 kV

The SEM images of the SiO$_2$ and Au QCM crystals coated with GO (Fig S2 a) and c)) show full coverage of the substrate with flakes, with the number of layers (determined from the contrast and further AFM) ranging from single to few layers overlaps, which is unavoidable when using spin coating deposition technique. The reduction of the GO (Fig S2 b) and d)) doesn't seem to affect the substrate coverage and the flakes density. However, in case of Au substrate (Fig S2 d)) the rGO flakes present many small holes (which is not an SEM artifact), unlike rGO present on SiO$_2$. Considering the identical reduction conditions for both samples, we speculate that the gas evolution during the GO reduction, combined with high temperature (180 °C) could have contributed to the Au etching which, in turn, contributed to the holes formation.

**AFM images**

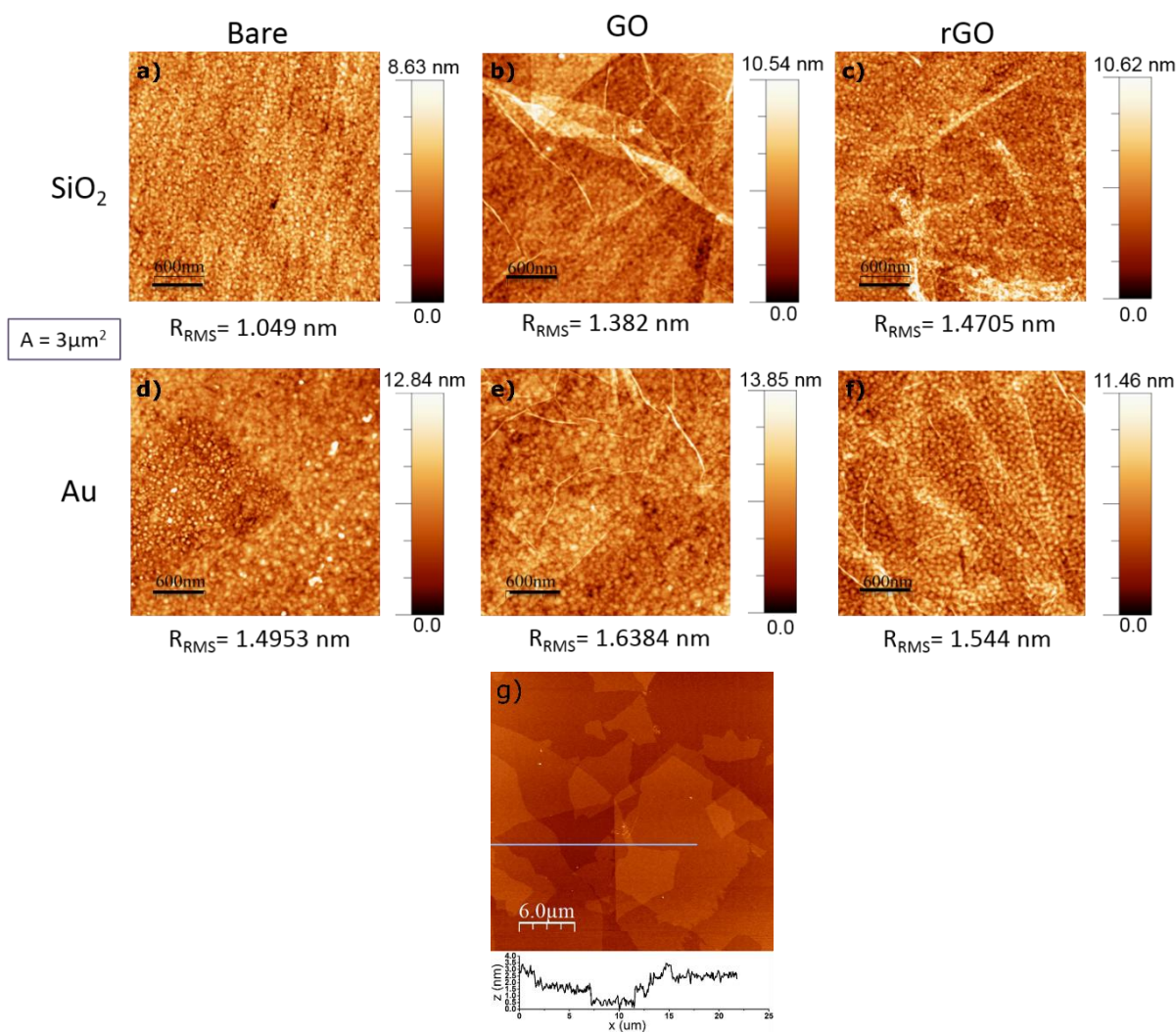

**Fig. S3** AFM mappings of full crystal set. Root-mean-square roughness ($R_{RMS}$) and height profile values in nm scale are shown for: **a)** bare $SiO_2$, **b)** GO-$SiO_2$, **c)** rGO-$SiO_2$, **d)** bare Au, **e)** GO-Au and **f)** rGO-Au. Scan area in all images is 3µm², **g)** shows the AFM image of GO reference sample on Si/SiO2 (290 nm) wafer for a 30/30 um surface scan. The AFM height profile shows a thickness of ~1nm for single layer flakes which increases almost proportionally with the number of flakes.

The topographical characterization of the prepared crystals was performed through Atomic Force Microscopy (AFM) to obtain a height profile and values for the root-mean-square roughness ($R_{RMS}$) for each crystal. The influence of the latter parameter on the response of the QCM has been stressed in different studies, comprising from the variation between a modeled frequency shift and experimental values of different RMS roughness levels (5), to the effect on the lipid-substrate interaction on the formation of structurally different areas of the same lipid composition (6). It has been shown that surface roughness affects the mechanisms of vesicle rupture and, in some cases, the formation of Supported Lipid Bilayers (SLBs) on solid supports (7), however SLB formation is only slightly affected on the nanometer scale. Therefore, controlling the roughness of a surface has direct impact on the structure of the membrane formed on top of the selected substrate. In fact, a rough crystal surface may effectively damp more the response of the frequency shift than a smooth polished crystal.

The AFM images for three individual $SiO_2$ and three Au crystals are presented in Fig S3. Each sample was carefully prepared by following the same steps and under similar conditions, as described in experimental section. Because the AFM scan has limited surface scan range, the information about the flakes distribution on the surface and the quality of the coverage are provided mainly be the SEM images. As the SEM showed, both, $SiO_2$ and Au QCM crystals are fully covered with GO/rGO with very few small empty spots, and from number of GO/rGO layers ranging mainly from single to 3 layers. The monolayer character of the original GO is confirmed by the reference sample (Fig S3 g)). However, it is difficult to ascribe in AFM the exact position or the number of layers present on the $SiO_2$ and Au QCMs substrates because of their high surface roughness (Fig S3 a) and d)) and the tendency of GO/rGO sheets to flatten on the surface and take its shape (Fig S3 b), c), e) and f)).

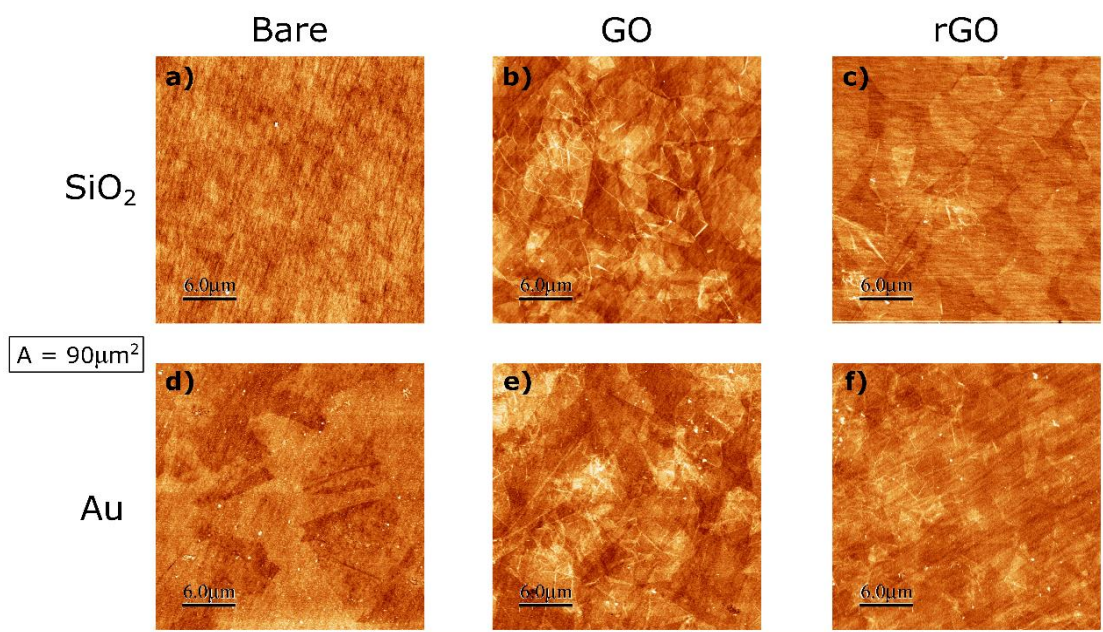

**Fig. S4** Large scan AFM mappings of full crystal set. **a)** bare SiO$_2$, **b)** GO-SiO$_2$, **c)** rGO-SiO$_2$, **d)** bare Au, **e)** GO-Au and **f)** rGO-Au. Scan area in all images is 90 μm$^2$.

The only reference of the flakes presence is detected by their crumbles, overlaps and creases formed during the spin coating and drying process, as can be seen clearly in lower resolution AFM images from Fig. S3.

As it can be seen from Fig S4 a), b) and c) the R$_{RMS}$ value of the surface increases with the addition of GO on the Si. A careful inspection of the Si-GO, however, shows that the roughness coming from the SiO$_2$ is slightly "smoothened" when the GO is present. This can be ascribed to the higher thickness of the flakes given by the functional groups and the water molecules trapped between the substrate and GO, between the GO flakes, and on the surface, due to the hydrophilic nature of GO. The increased R$_{RMS}$ value is probably given by the contribution of the wrinkles, folds and overlaps of GO flakes to the existing roughness. In case of Si-rGO substrate, the roughness of the substrate seems very similar to the bare Si. An explanation would be the reduction in thickness of the GO flakes upon the reduction process accompanied by the dehydration. These, together with the wrinkled nature of the rGO flakes, will contribute to a higher R$_{RMS}$ value compared to bare Si and GO.

The intrinsic higher roughness of the Au substrate (Fig S4 d)) doesn't change significantly with the addition of GO (Fig S4 e)). Unlike the case of SiO$_2$, in this case the GO coated Au seems to keep the roughness characteristics and the only contribution to the slightly increased R$_{RMS}$ value is the roughness generated by the flakes, at, however, lower rate than in case of SiO$_2$. This can be due to the difference in GO – substrate interaction, as well as more hydrophobic nature of Au which leads to a better dehydration between GO and substrate. After the thermal reduction, the R$_{RMS}$ values for the Au-rGO (Fig S4 f)) are lower than Au-GO and slightly higher than bare Au. A close look at the AFM scan (Fig. S4 f)) reveals that the deposited Au "islands" present on Au-rGO have a more flat and uniform character compared to the initial Au substrate. This can be due to a slight Au etching during the high temperature reduction of GO, which would explain lower roughness compared to Au-GO sample.

**XPS on GO and rGO coated substrates**

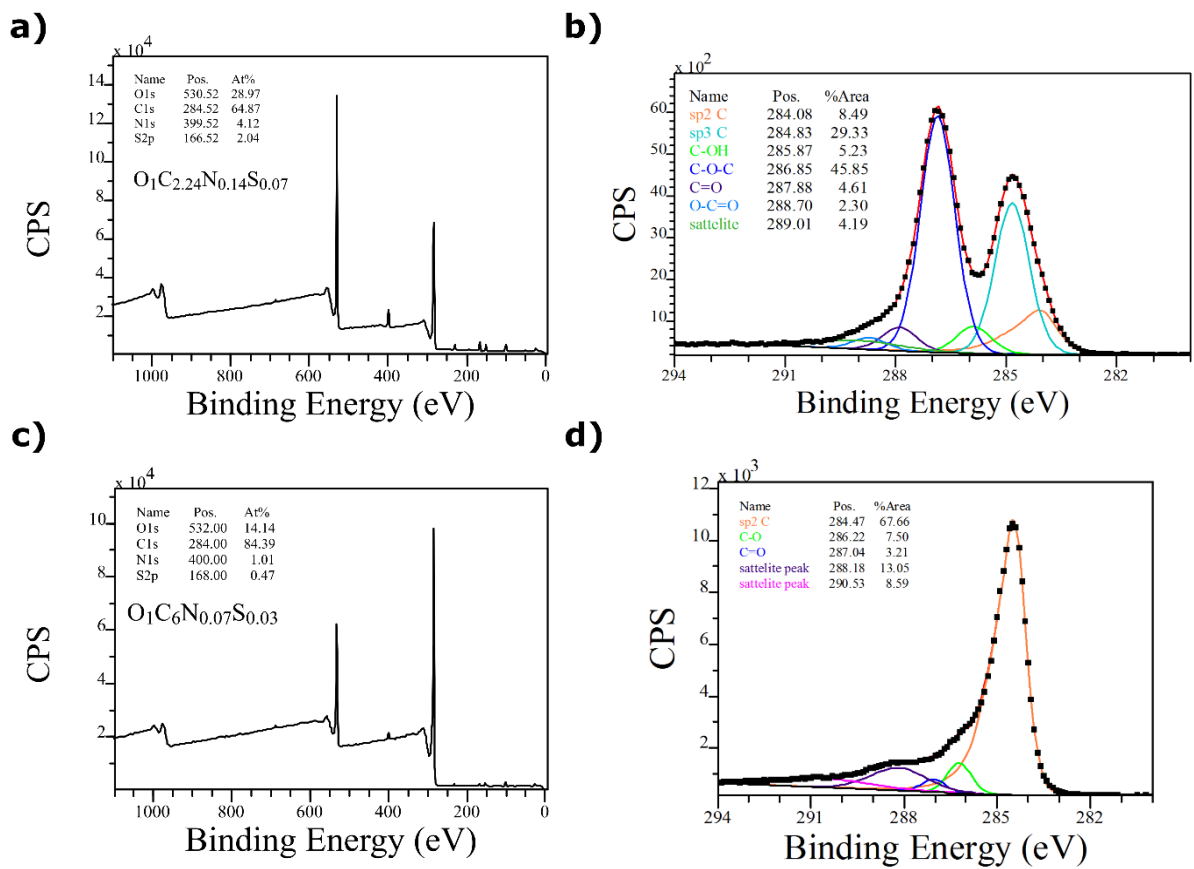

**Fig. S5** XPS of coated samples **a)** wide scan GO, **b)** deconvoluted C1s of GO, **c)** wide scan rGO, **d)** deconvoluted C1s of rGO.

The XPS technique was performed to reveal the nature of chemical bonds in GO and to monitor their evolution after GO reduction. Fig S5 a), b), c) and d) represent the wide scan and C1s spectra of GO and rGO respectively. The wide scan of GO reveals a C to O ratio of ~2, in accordance with the literature for GO (8) with small amounts of nitrogen and Sulphur impurities. After reduction (Fig. S5 c)) the C to O ratio increases significantly to 6. The C1s spectrum of the GO (Fig. S5 b) shows the presence of different functional groups decorating the basal plane and the edges of GO: hydroxyl (C-OH) and epoxy (C-O-C) groups between ~285 and 287 eV, carbonyl (C=O) and carboxyl (O-C=O) groups between ~287 and 289 eV, and finally, $sp^2$ and $sp^3$ carbons – around ~284 eV. After reduction (Fig S5 d)) the rGO presents fewer oxygen groups, i.e. single and double carbon –oxygen groups with a binding energy of ~286 and 287 eV respectively, and an increased intensity $sp^2$ carbon peak. This proves the reduction of GO to rGO and a significant restoration of $sp^2$ carbons.

**Raman mappings**

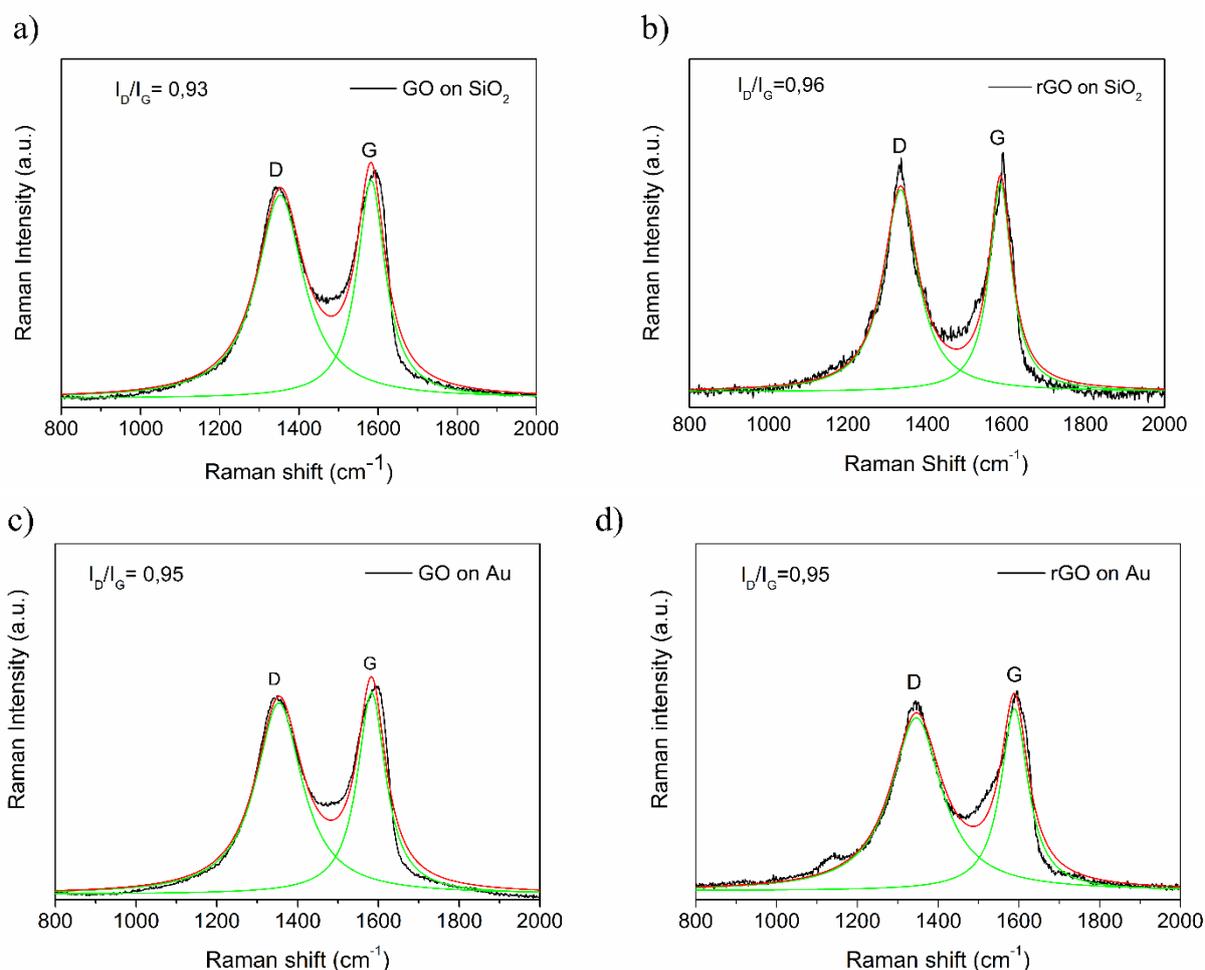

**Fig S6**. Raman spectra (with fits for G and D peak components) of (a) $SiO_2$ surface with GO coating, (b) $SiO_2$ surface with rGO coating, (c) Au surface with GO coating and (d) Au surface with rGO coating. Green curves show peak fits and the red curves show the sum of the peak fits (color online version).

Raman is a powerful technique used for the characterization of the graphitic materials, providing information about number of layers, lattice defects, doping etc. (9,10). Fig S6 shows the Raman spectra of GO and rGO coated QCM sensors prepared as described before. Peak fit is shown on the curves in green (online version). One of the spectral features of graphene is associated with the optical phonon mode, which occurs around ~ 1580 $cm^{-1}$ and is called the G band (10). The D peak is associated with defects in the structure ($sp^3$ bonding) appears at ~ 1350 $cm^{-1}$ (11). The relative intensity of D to G provides an indicator for determining the in-plane crystallite size or the amount of disorder in the sample, indicating the $sp^2/sp^3$ carbon ratio, ergo, it shows the disorder or the restoration of the graphene lattice (9,12).

Figures S6 a) and b) show the Raman spectra of GO and rGO on $SiO_2$-QCM-D sensors, respectively. The $I_D/I_G$ value of 0.96 suggests the presence of graphitic domains after the reduction process in $SiO_2$ (Fig. S6 b)) while the ratio obtained for GO is equal to 0.93 (Fig. S6 a)). Similarly, Figures S6 c) and d) show the Raman spectra for GO and rGO, respectively, on Au-QCM-D sensors. The $I_D/I_G$ ratio on Fig. S6 c) and d) remains equal according to our data fit, suggesting equivalent defectiveness and the absence of any damage due to the reduction process on the scanned regions. Overall, these Raman spectra indicates the presence of graphene and graphene-like domains on the selected substrates.

# Overall frequency and dissipation values (3$^{rd}$, 5$^{th}$ & 7$^{th}$ harmonics)

**Bare substrates**

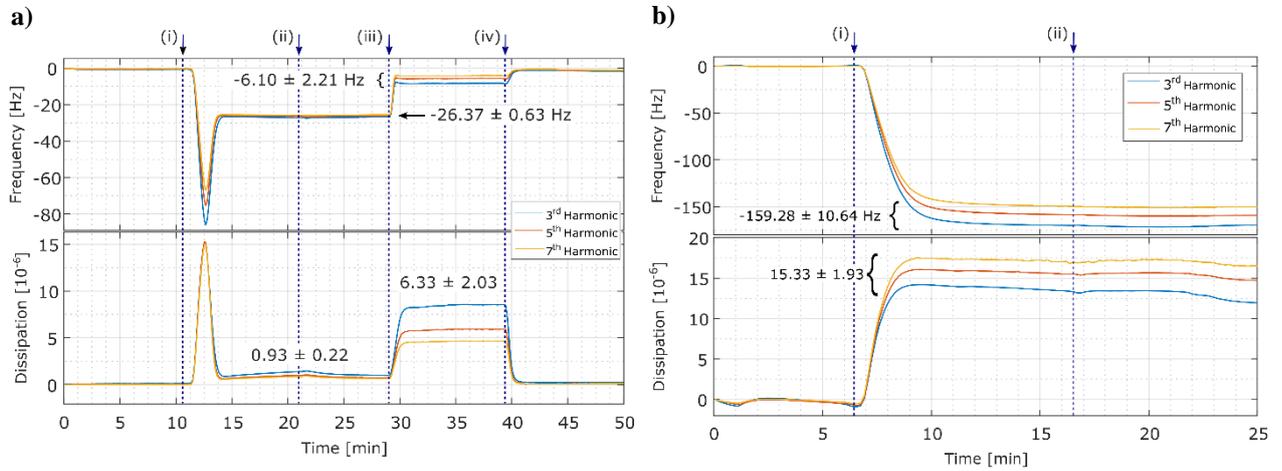

**Fig.S7** Adsorption of DOPC **a)** on bare SiO$_2$ **b)** on bare Au. Steps: (i) injection of DOPC then (ii) buffer rinse, (iii) SDS wash, (iv) final buffer rinse.

**GO coated substrates**

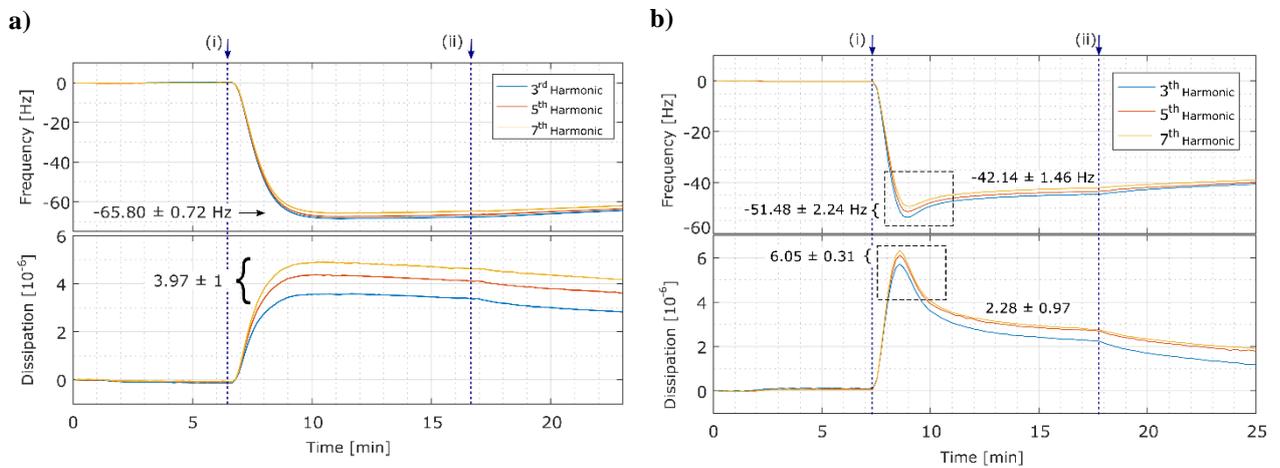

**Fig.S8** Adsorption of DOPC **a)** on GO-SiO$_2$ **b)** on GO-Au. Steps: (i) injection of DOPC then (ii) buffer rinse

**rGO coated substrates**

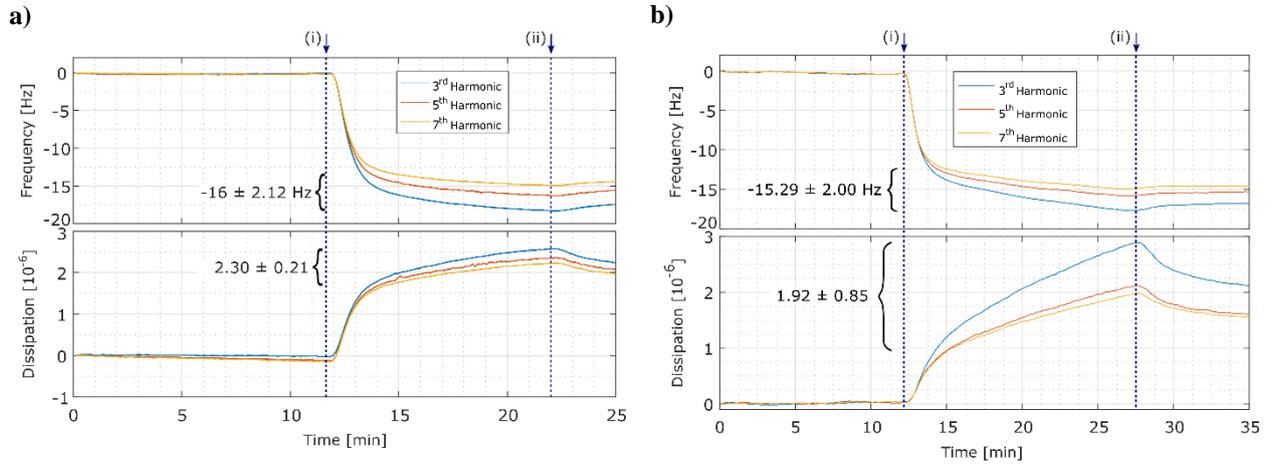

**Fig.S9** Adsorption of DOPC **a)** on rGO-SiO$_2$ **b)** on rGO-Au. Steps: (i) injection of DOPC then (ii) buffer rinse

**Binding event on bare substrates**

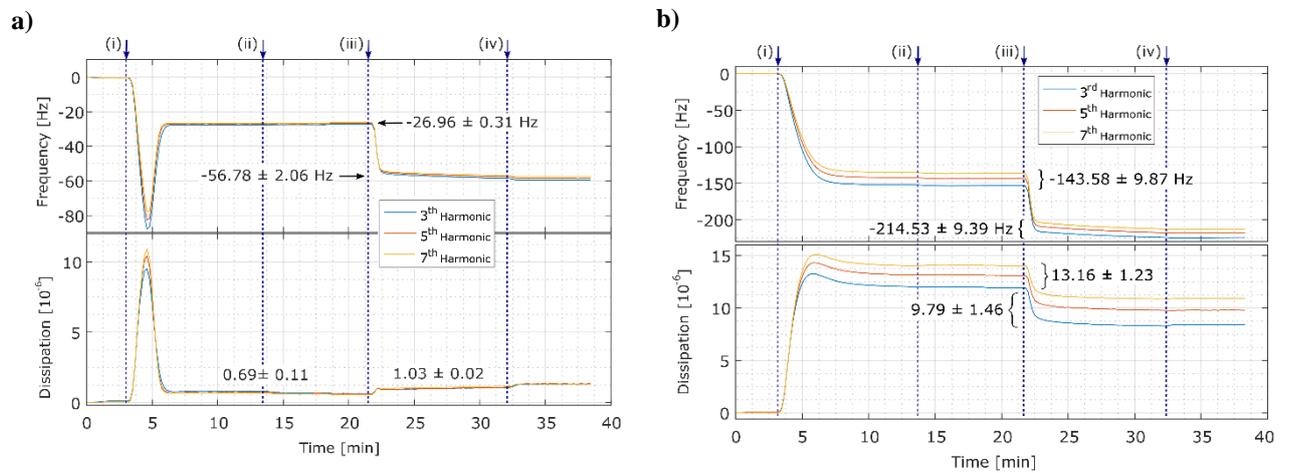

**Fig.S10** Biotin-Avidin binding event on **a)** a lipid bilayer on SiO$_2$ **b)** intact vesicles on Au. Steps: (i) injection of DOPC then (ii) buffer rinse (iii) Avidin injection, (iv) final buffer rinse

## Binding event on rGO-coated substrates

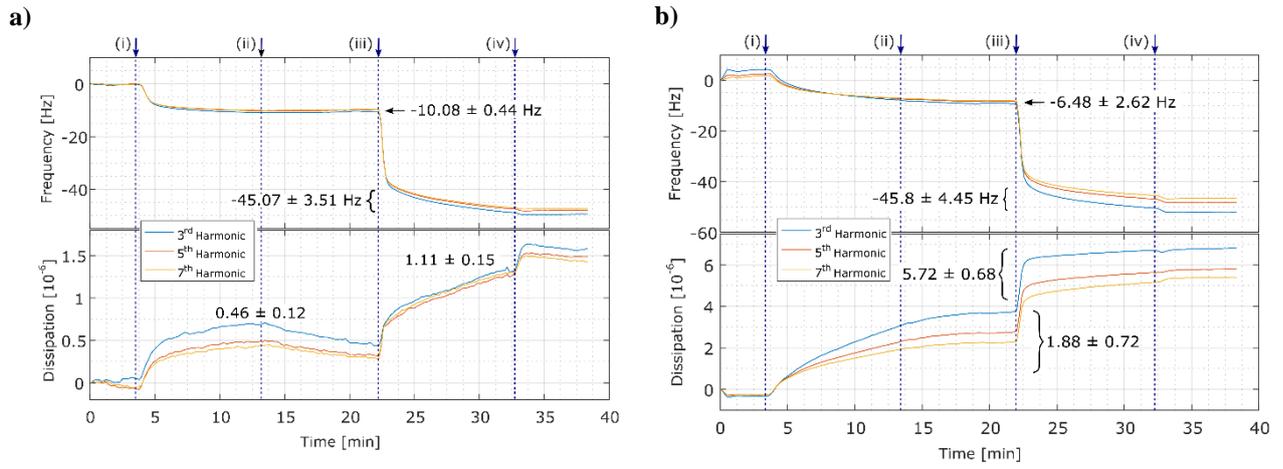

**Fig.S11** Biotin-Avidin binding event on **a)** a lipid monolayer on rGO-SiO$_2$ **b)** a lipid monolayer on rGO-Au. Steps: (i) injection of DOPC then (ii) buffer rinse (iii) Avidin injection, (iv) final buffer rinse